\documentclass[aps,twocolumn,showpacs,preprintnumbers,amsmath,amssymb,nofootinbib,superscriptaddress,notitlepage]{revtex4-1}

\usepackage{epsfig}

\begin{document}

\title{Deriving the existence of $B\bar{B}^*$ bound states from the $X(3872)$
and Heavy Quark Symmetry}

	\author{J. Nieves}
\affiliation{Departamento de F\'{\i}sica Te\'orica and
Instituto de F\'{\i}sica Corpuscular (IFIC), Centro Mixto CSIC-Universidad de Valencia, Institutos de Investigaci\'on de Paterna, Aptd. 22085, E-46071 Valencia, Spain}
	\author{M. Pav\'on Valderrama}\email{m.pavon.valderrama@ific.uv.es} 
\affiliation{Departamento de F\'{\i}sica Te\'orica and
Instituto de F\'{\i}sica Corpuscular (IFIC), Centro Mixto CSIC-Universidad de Valencia, Institutos de Investigaci\'on de Paterna, Aptd. 22085, E-46071 Valencia, Spain}



\begin{abstract} 
\rule{0ex}{3ex}
We discuss the possibility and the description of bound states
between $B$ and $\bar{B}^*$ mesons.
We argue that the existence of such a bound state can be deduced
from (i) the weakly bound $X(3872)$ state, (ii) certain assumptions
about the short range dynamics of the $D\bar{D}^*$ system and
(iii) heavy quark symmetry.
From these assumptions the binding energy of the possible $B\bar{B}^*$
bound states is determined, first in a theory containing only contact
interactions which serves as a straightforward illustration of the method,
and then the effects of including the one pion exchange potential
are discussed.
In this latter case three isoscalar states are predicted:
a positive and negative C-parity $^3S_1-{}^3D_1$ state
with a binding energy of $20\,{\rm MeV}$ and $6\,{\rm MeV}$ below threshold
respectively, and a positive C-parity $^3P_0$ shallow state located almost at
the $B\bar{B}^*$ threshold.
However, large uncertainties are generated as a consequence of the $1/m_Q$
corrections from heavy quark symmetry.
Finally, 
the newly discovered isovector $Z_b(10610)$ state can be easily accommodated
within the present framework by a minor modification of the short
range dynamics.
\end{abstract}


\maketitle

\section{Introduction}

The existence of heavy meson bound states, a possibility first theorized
by Voloshin and Okun~\cite{Voloshin:1976ap},
is grounded on the observation that meson-exchange forces can arise
between two heavy mesons as a consequence of their light quark content.
In analogy with 
the nuclear forces in the two-nucleon system,
these exchange forces may eventually be strong enough as to generate
bound states and resonances in two-meson systems.
In particular, we expect the one pion exchange (OPE) potential 
to drive the long range behaviour of
two-meson systems and therefore to play a very special role
in the formation and the description of
their corresponding bound states.
The seminal works of T\"ornqvist~\cite{Tornqvist:1991ks,Tornqvist:1993ng},
Manohar and Wise~\cite{Manohar:1992nd} and
Ericson and Karl~\cite{Ericson:1993wy},
which already stressed the importance of pion exchanges in the formation of
deuteron-like meson bound states,
indicated that these states are more probable (i.e. more bound)
in the bottom sector than in the charm one.

The discovery of the $X(3872)$ state by the Belle
collaboration~\cite{Choi:2003ue}, later confirmed by  CDF~\cite{Acosta:2003zx},
D0~\cite{Abazov:2004kp} and BABAR~\cite{Aubert:2004ns}, probably
represents the most obvious candidate for a heavy meson bound state.
With a mass of $m_X = 3871.56 \pm 0.22\,{\rm MeV}$~\cite{Nakamura:2010zzi}
the $X(3872)$ lies very close to the $D^0\bar{D}^{*0}$ threshold,
$m_{D^0} + m_{D^{*0}} = 3871.79 \pm 0.21 \,{\rm MeV}$,
suggesting the natural interpretation of a shallow bound or virtual state.
The average separation between the $D^0$ and $D^{*0}$ mesons,
$\sqrt{\langle r^2  \rangle} \sim 10\,{\rm fm }$,
will enhance strongly the role of the molecular component at low energies
in comparison with the more compact multiquark components.
It should be noted however that the molecular interpretation is only possible
if the quantum numbers of the $X(3872)$ are $J^{PC} = 1^{++}$,
a value compatible with available experimental
information~\cite{Abe:2005iya,Abulencia:2006ma}.
In this regard, the new experimental analysis
of Ref.~\cite{delAmoSanchez:2010jr}, which prefers the quantum numbers $2^{-+}$,
raises the likelihood of the tetraquark and charmonium interpretations over
the molecular hypothesis.
On a related note, the recent finding of the $Z_b(10610)$ and $Z_b(10650)$
states by the Belle collaboration~\cite{Collaboration:2011gj},
very close to the $B\bar{B}^*$ and $B^*\bar{B}^*$ thresholds
respectively, might also prove to be heavy meson bound states. 

The molecular descriptions of the $X(3872)$ can be classified in
two broad types: (i) the short-range interaction
picture~\cite{Close:2003sg,Voloshin:2003nt,Braaten:2003he,Voloshin:2005rt,AlFiky:2005jd,Gamermann:2009fv,Gamermann:2009uq},
in which the $X(3872)$ is the result of contact interactions
between the $D$ and $\bar{D}^*$ mesons,
and (ii) the potential picture~\cite{Suzuki:2005ha,Thomas:2008ja,Liu:2008fh,Liu:2008tn,Lee:2009hy,Ding:2009vj,Xu:2010fc},
in which the $X(3872)$ is bound as the result of meson-exchanges.
A middle path is provided by X-EFT~\cite{Fleming:2007rp},
in which the $X(3872)$ is bound due to the effect of the short range dynamics,
but pions are included as perturbations.

The physical motivation behind the short-range description of the $X(3872)$
is that the two particle conforming a low-lying bound state are too far apart
from each other as to distinguish the specific details of the interaction
binding them, a phenomenon called
{\it universality}~\cite{Braaten:2003he}.
However, although universality demystifies the success of contact theories,
there are observables which cannot be explained without the explicitly
consideration of shorter range components of the $X(3872)$.
A clarifying example is the large branching ratio~\cite{Abe:2005iya}, 
\begin{eqnarray}
\frac{Br(X \to J/\psi \pi^{+} \pi^{-} \pi^0)}
{Br(X \to J/\psi \pi^{+} \pi^{-})} = 1.0 \pm 0.4 \pm 0.3\, ,
\end{eqnarray}
which finds a most natural explanation by considering the charged
$D^+ D^{*-}$ component
in the $X(3872)$~\cite{Gamermann:2009fv,Gamermann:2009uq}.

The potential description of the $X(3872)$, which was pioneered by
T\"ornqvist~\cite{Tornqvist:1993ng}, 
is constructed in analogy with the traditional meson
theory of the nuclear forces~\cite{Machleidt:1987hj,Machleidt:1989tm}.
In this picture, the $D$ and $\bar{D}^*$ mesons interact
via a series of exchanges of increasingly heavy mesons,
conforming a long range potential.
The pion is always included, while more sophisticated descriptions
also consider the effect of two pion~\cite{Xu:2010fc}
and heavier meson exchanges~\cite{Liu:2008tn,Lee:2009hy}.
However, pseudoscalar and vector meson exchanges give rise
to a singular tensor interaction, which diverges as $1/r^3$
at short enough distances.
This singularity requires the regularization of the potential,
usually by the inclusion of a form factor.
Usually, these approaches do not only consider the $D\bar{D}^*$ system,
but also other heavy meson systems, in particular $B\bar{B}^*$.

The conclusions which are derived from the potential picture are
easy to summarize: in first place, meson exchange forces,
specially the pion, are too weak to be able to bind the
$D$ and $\bar{D}^*$ mesons together (unless the cut-off
in the form factor takes very large values);
second, if the argument is extended to the $B$ and $\bar{B}^*$
meson system, the conclusion is that they are probably bound.
However, it is very difficult to determine the binding energy of
this two-body system.
This is just a restatement of the argument of
Ericson and Karl~\cite{Ericson:1993wy}:
a system with a higher reduced mass is more likely to bind,
as a result of the lower kinetic energy,
meaning that the existence of a bound state between B mesons
is highly likely.

In this work we employ two different effective field theory (EFT) descriptions
of heavy meson systems at lowest order: (i) a pionless theory,
in which the mesons interact through local contact interactions,
and (ii) a pionfull theory, in which we include the pion exchanges explicitly
(for reviews, see~\cite{Beane:2000fx,Bedaque:2002mn,Epelbaum:2005pn,Epelbaum:2008ga}).
The EFT approach we follow is cut-off EFT~\cite{Lepage:1997cs,Lepage:1989hf},
in which the potential is expanded according to naive dimensional analysis
\begin{eqnarray}
V_{\rm EFT}(\vec{q}) = 
V^{(0)}(\vec{q}) + V^{(2)}(\vec{q}) + \mathcal{O}({P^3}) \, ,
\end{eqnarray}
where $P$ represents the low energy scales of the system,
in this case the pion mass and the relative momenta
of the heavy mesons.
Finally, a cut-off is included in the computation
to regularize the ultraviolet divergences.
The resulting potential is iterated in the Schr\"odinger or Lippmann-Schwinger
equation in order to compute the observables.
If pions are included, the approach basically coincides with
the Weinberg counting~\cite{Weinberg:1990rz,Weinberg:1991um}
as applied in the two nucleon system, which is known
to be very successful phenomenologically~\cite{Entem:2003ft,Epelbaum:2004fk}.
The optimal cut-off choice is to be taken of the order of the high energy
scale of the system~\cite{Epelbaum:2009sd},
that is, the scale at which the short range effects
not explicitly taken into account in the EFT start to manifest.
This corresponds, for example, with the energy scale at which the composite
nature of the heavy mesons can be probed or for which the exchange
of heavier mesons (like the $\rho$) needs to be taken into account,
suggesting a value for the hard scale of
$\Lambda_0 \sim 0.5-1\,{\rm GeV}$.
In this work, the natural value of cut-off will be roughly interpreted
as the inverse size of the heavy mesons.

As a consequence of the scarce phenomenological input available
about the $D\bar{D}^*$ and $B\bar{B}^*$ interactions,
we limit ourselves to a leading order (${\rm LO}$ or order $P^0$) EFT,
in which there is only one low energy contact operator ($C_0$).
The long range interaction at ${\rm LO}$ only contains the one pion exchange
potential, resulting in a very compact and simple description
of the two meson system.
In orthodox implementations of the EFT concept, the counterterm is a running
coupling constant $C_0 = C_0 (\Lambda)$, which can be determined
for a given cut-off $\Lambda$ by reproducing, for example,
the binding energy of the system (or any other observable).
However the previous idea, although well suited for the $D\bar{D}^*$ case,
cannot be employed in the $B\bar{B}^*$ system as there is no
experimental information available, posing a problem
for the present EFT formulation.

The alternative we follow is the approach of Ref.~\cite{Gamermann:2009uq}:
we assume the counterterm $C_0$ to be saturated by short-range dynamics,
while the cut-off is determined by reproducing the binding energy
in the $D\bar{D}^*$ system.
The extension of the previous scheme to the bottom sector is straightforward
once we consider heavy quark symmetry (HQS)~\cite{Isgur:1989vq,Isgur:1989ed,Neubert:1993mb,Manohar:2000dt},
which allows us (i) to extrapolate the saturation condition
for the counterterm from
the $D\bar{D}^*$ to the $B\bar{B}^*$ system and (ii) to expect
the value of the cut-off to be similar in both cases.
With these two pieces of information we will be able to make
predictions in the bottom sector.
This approach, which can be labelled EFT {\it mapping}
or {\it saturation} method, does not correspond
to the standard formulation of the EFT concept.
However, if the determination of the cut-off results in a natural value,
the saturation condition will not contradict either any of
the usual EFT tenets.
The fact is that the saturation of the low energy constants
by short-range physics is a sensible prospect within
the EFT context, as exemplified in Ref.~\cite{Epelbaum:2001fm}
for the two-nucleon sector.
The important point is that saturation and HQS allow us
to correlate the charm and bottom sectors.

The manuscript is organized as follows:
in Sect.\ref{sec:formalism} we present the formalism to describe bound states
between heavy mesons, in Sect.\ref{sec:contact} we show the prediction for
a $B\bar{B}^*$ bound state based on a contact (i.e. pionless) theory,
which serves as a straightforward exposition of our approach.
In Sect.\ref{sec:OPE} we discuss the role played by the inclusion
of the one pion exchange potential, which implies
the appearance of a new $B\bar{B}^*$ P-wave
bound state.
Finally, in Sect.\ref{sec:discussion} we present our conclusions.

\section{Formalism}
\label{sec:formalism}

In this section we will describe the two heavy meson system
as a non-relativistic two-body problem in quantum mechanics,
for which the potential between the heavy mesons is
a well-defined object.
We formulate the bound state problem in momentum space,
where the wave functions are obtained by solving
the Lippmann-Schwinger equation.
We also consider in detail the explicit treatment of tensor forces,
which will appear as a consequence of pseudoscalar meson exchange,
and coupled channels.

\subsection{Bound State Equation}

The Lippmann-Schwinger equation for a bound state reads
\begin{eqnarray}
\label{eq:BS}
| \Psi_B \rangle = G_0(E_B) V | \Psi_B \rangle \, ,
\end{eqnarray}
where $| \Psi_B \rangle$ represents the wave function of the bound state
system, $V$ the two-body potential and $G_0(E)$ is the resolvent operator,
which is given by
\begin{eqnarray}
G_0(E) = \frac{1}{E - H_0} \, ,
\end{eqnarray}
where $E < 0$ ($ > 0$) for bound (scattering) states.
In the expression above, $H_0$ is the free Hamiltonian of
the two particle system, which is defined as
\begin{eqnarray}
H_0 |\vec{k}\rangle = \frac{\vec{k}^2}{2\mu}\,|\vec{k}\rangle \, ,
\end{eqnarray}
where the center-of-mass motion has been removed,
$\mu$ is the reduced mass of the system and $| \vec{k} \rangle$
represents a two-particle state with relative momentum $\vec{k}$
in the center-of-mass system.
Projecting the previous equation onto plane waves,
we get an explicit representation of the bound state equation
\begin{eqnarray}
\Psi_B(\vec{k})
=&& \nonumber \\ - \frac{2\mu}{k^2 + \gamma^2} &&
\int \frac{d^3\vec{k}\,'}{(2\pi)^3}\,
\langle \vec{k}| V | \vec{k}\,' \rangle 
\, \Psi_B(\vec{k}\,') \, , 
\end{eqnarray}
with $\Psi_B(\vec{k}) = \langle \vec{k} | \Psi_B \rangle$,
$\gamma^2 = -2\mu\,E_B$ and where $E_B$($< 0$)
is the bound state energy.
Finally, the solution is subjected to the normalization condition
\begin{eqnarray}
\langle \Psi_B | \Psi_B \rangle = \int \frac{d^3\vec{k}}{(2\pi)^3}\,
|\Psi_{B} (\vec{k})|^2 = 1 \, .
\end{eqnarray}

\subsection{Central Potential}

If we assume a central potential, the previous equation can be projected
onto the partial wave basis, defined as
\begin{eqnarray}
| \vec{k} \rangle = \sqrt{4\pi}
\sum_{l m} | k, l m \rangle \, Y_{l m}^{(*)}(\hat{k}) \, ,
\end{eqnarray}
from which we can expand the bound state wave function as follows
\begin{eqnarray}
\label{eq:PsiB_l}
\Psi_B(\vec{k}) = \sqrt{4\pi}
\sum_{l m} \langle k, l m | \Psi_B \rangle \, Y_{l m}(\hat{k}) \, .
\end{eqnarray}
For a central potential the angular momentum of the two-body system
is conserved and the matrix elements of the potential
in the $| k, l m \rangle$ basis fulfill
the relation
\begin{eqnarray}
\label{eq:V_l}
\langle k, l m | V | k', l' m' \rangle = 
\langle k | V_l| k' \rangle \delta_{l l'} \delta_{m m'} \, ,
\end{eqnarray}
in which $\delta_{l l'}$ and $\delta_{m m'}$ take into account
the fact that the potential is spherically symmetric.
The relationship between the projected potential and the potential
in the plane wave basis is given by
\begin{eqnarray}
\langle k | V_l| k' \rangle = \frac{1}{4\pi}\int d\hat{k}\,d\hat{k}\,'
Y_{lm}^{(*)}(\hat{k})\,\langle \vec{k} | V | \vec{k}\,' \rangle
Y_{l m}(\hat{k}')\, ,
\end{eqnarray}
where the result does not depend on the third component of
the angular momentum $m$.

For a central potential the bound state has well-defined quantum numbers
$l$ and $m$ and the partial wave expansion of $\Psi_B(\vec{k})$,
Eq.~(\ref{eq:PsiB_l}), simplifies to
\begin{eqnarray}
\Psi_{B (l m)}(\vec{k}) = \sqrt{4\pi}\,\Psi_{B,l}(k) \, Y_{l m} (\hat{k}) \, ,
\end{eqnarray}
where the partial wave projected wave function $\Psi_{B,l}(k)$ does
not depends on $m$ as a consequence of Eq.~(\ref{eq:V_l}).
In such a case, the bound state equation simplifies to
\begin{eqnarray}
\label{eq:bs-central}
\Psi_{B, l} (k)
=&& \nonumber \\ - \frac{2\mu}{k^2 + \gamma^2} &&
\int \frac{k'^2 dk'}{2\pi^2}\,
\langle k | V_{l}| k' \rangle \Psi_{B, l} (k') \, , 
\end{eqnarray}
where the partial wave projected wave function obeys
the normalization condition
\begin{eqnarray}
\int \frac{k^2 dk}{2\pi^2}\,|\Psi_{B, l} (k)|^2 = 1 \, . 
\end{eqnarray}

\subsection{Tensor Forces}
\label{sec:tensor_forces}

The longest range term of the interaction between a heavy pseudoscalar and
vector meson (both of which contain a light quark) is due to one pion exchange,
which in turn implies the existence of a tensor component in the potential.
This tensor force will mix channels with different angular momenta.
However, the total angular momentum and the parity of
the two meson system will be conserved.
Therefore, instead of projecting into the $| k, l m \rangle$ partial waves,
we will rather consider the following states
\begin{eqnarray}
| k, l j m \rangle = \sum_{m_l, m_s} |k, l m_l \rangle \, | 1 \lambda \rangle
\,\langle 1 \lambda l m_l | j m \rangle \, ,
\end{eqnarray}
where $| 1 \lambda \rangle$ is the intrinsic spin state of the 
the pseudoscalar-vector meson system (which can be identified
with the polarization vector of the vector meson),
and $\langle 1 \lambda l m_l | j m \rangle$
is a Clebsh-Gordan coefficient.
In the previous basis, the partial wave expansion of the plane wave reads
\begin{eqnarray}
| \vec{k}, 1 \lambda \rangle = \sqrt{4\pi}
\sum_{l j m} | k, l j m \rangle \,
{\mathcal{Z}^{l \lambda}_{j m}\,}^{(*)}(\hat{k}) \, ,
\end{eqnarray}
where the sum over $l$ runs from $| j - 1 |$ to $| j + 1 |$, and with
\begin{eqnarray}
\mathcal{Z}^{l \lambda}_{j m}(\hat{k}) =
\langle 1 \lambda l m_l | j m \rangle\,Y_{l m_l}(\hat{k}) \, .
\end{eqnarray}
Conservation of the total angular momentum implies that the matrix elements
of the potential are diagonal in $j$ and $m$, that is
\begin{eqnarray}
\label{eq:tensor-potential-projection}
\langle k, l j m | V | k', l' j' m' \rangle = 
\langle k | V_{l l' j}| k' \rangle \delta_{j j'} \delta_{m m'} \, .
\end{eqnarray}
In addition, as a consequence of parity conservation, the matrix elements
between odd $l$ and even $l'$, and viceversa, are zero.
The relationship between the projected and unprojected potential reads
\begin{eqnarray}
\label{eq:V_llj}
\langle k | V_{l l' j}| k' \rangle = \frac{1}{4\pi} && \sum_{\lambda, \lambda'}\,
\int d\hat{k}\,d\hat{k}\,'{{\mathcal Z}^{l \lambda}_{jm}}^{(*)}(\hat{k})\,
\nonumber \\
&\times& \langle \vec{k}, 1 \lambda | V | 
\vec{k}\,' , 1 \lambda' \rangle \,
{\mathcal Z}^{l' \lambda'}_{jm}(\hat{k}')\, , \nonumber \\
\end{eqnarray}
where we have taken into account that the unprojected potential between
the pseudoscalar and vector mesons depends on the polarizations
$\lambda$ and $\lambda'$ of the initial and final states.

Owing to the properties of the tensor force, the bound state between a
pseudoscalar and a vector meson has well-defined total angular momentum
and parity.
In such a case, the partial wave projection of the bound state wave function
can be written as
\begin{eqnarray}
\label{eq:Psi-p-partial-wave}
\Psi_{B (p j m)}(\vec{k}) = \sqrt{4\pi}\,
\sum_{{\{l\}}_p}\,\Psi_{B,l j}(k) \,\mathcal{Y}_{l j m}(\hat{k})
\end{eqnarray}
with $\Psi_{B,l j}$ the projected wave function, and where
the notation ${\{l\}}_p$ is used to denote the set of angular
momenta with parity $p$, and with
\begin{eqnarray}
\mathcal{Y}_{l j m}(\hat{k}) =
\sum_{\lambda} \mathcal{Z}^{l \lambda}_{j m}(\hat{k})\,| 1 \lambda \rangle \, .
\end{eqnarray}
It should be noted that the projected wave function $\Psi_{B,l j}$
does not depend on the third component of the total angular
momentum ($m$) as a consequence of Eq.~(\ref{eq:tensor-potential-projection}).
The partial wave projection of the bound state equation reads in this case as
\begin{eqnarray}
\label{eq:bs-eq-tensor-projection}
\Psi_{B, l j} (k)
=&& \nonumber \\ - \frac{2\mu}{k^2 + \gamma^2} &&
\sum_{l'} \int \frac{k'^2 dk'}{2\pi^2}\,
\langle k | V_{l l' j}| k' \rangle \Psi_{B, l' j} (k') \, , 
\end{eqnarray}
As can be seen, parity conservation implies that channels with $l = j$
are uncoupled while channels with $l = j \pm 1$ are coupled.
Finally, the normalization condition for the wave function reads
\begin{eqnarray}
\sum_{{\{l\}}_p}\,\int \frac{k^2 dk}{2\pi^2}\,
|\Psi_{B, l j} (k)|^2 = 1 \, .
\end{eqnarray}

In general we will prefer the spectroscopic notation $^{2S+1}L_J$
to the $(pjm)$ notation in this work.
In the spectroscopic notation,
$S$ refers to the total intrinsic spin of the two body
system, which for the particular case of a $D$ and $\bar{D}^*$ mesons
is always $S=1$, $L$ is the orbital angular momentum and $J$ is the total
angular momentum.
For coupled angular momentum channels, we will employ a dash to indicate
the different angular momentum components.
For example, the $p = +1$, $j = 1$ state, which couples the S- and D-waves,
will be denoted by $^3S_1-{}^3D_1$; on the contrary, the $p = -1$, $j=1$
state, which only contains a P-wave, will be simply the $^3P_1$ channel
in this notation.

\subsection{Coupled Channels}
 
In certain cases we will need to consider the existence of different channels
in the description of heavy meson bound states,
mainly due to isospin breaking effects.
For example, the $X(3872)$ contains a neutral and a charged component
\begin{eqnarray}
\label{eq:X-nc}
| X(3872) \rangle &=& 
\frac{1}{\sqrt{2}}\,
\left[ 
| D^0 \bar{D}^{*0} \rangle - | D^{*0} \bar{D}^0  \rangle 
\right]\,| X_0 \rangle \nonumber \\
&+& 
\frac{1}{\sqrt{2}}\,
\left[ 
| D^{+} {D}^{*-} \rangle - | D^{*+} {D}^{-} \rangle 
\right]\,|X_C \rangle \, , \nonumber \\
\end{eqnarray}
where, if isospin were conserved, we would have
$|X_0 \rangle = |X_C \rangle$ for the isoscalar case ($I=0$).
The previous representation assumes that the $X(3872)$
is a positive C-parity state~\footnote{We remind that C-parity is
a good quantum number for a meson-antimeson system. We are also
following the negative C-parity convention for the vector
meson, in which $\hat{C} | D^* \rangle = - | \bar{D}^* \rangle$,
which is the most natural one.}.
Owing to isospin breaking effects, the interaction can vary slightly
depending on the channel and in addition the channels can have
different kinematic thresholds which need to be
taken into account.

The extension to the coupled-channel case is trivial, and only requires
to consider the existence of different components in the wave function
\begin{eqnarray}
| \Psi_B \rangle = \sum_{\alpha} | \Psi_B^{\alpha} \rangle \, ,
\end{eqnarray}
where we use upper indices to denote the different channels
(lower indices are reserved for angular momentum).
For the previous wave function, the bound state equation reads
\begin{eqnarray}
| \Psi_B^\alpha \rangle = G_{0}^{\alpha}(E_B) \sum_{\beta} V^{\alpha \beta} 
| \Psi_B^\beta \rangle \, ,
\end{eqnarray}
where the potential $V^{\alpha \beta}$ is now a matrix,
and $G_{0}^{\alpha}(E)$ takes the form
\begin{eqnarray}
G_0^{\alpha}(E) = \frac{1}{E - H_{0}^{\alpha}} \, ,
\end{eqnarray}
with $H_0^{\alpha}$ the eigenvalue of the free Hamiltonian operator
in the plane wave basis for the $\alpha$ channel
\begin{eqnarray}
H_0 |\vec{k},\alpha \rangle = H_0^{\alpha} |\vec{k},\alpha \rangle =
\left[ \frac{k^2}{2\mu_{\alpha}} - \Delta_{\alpha} \right]\,
|\vec{k},\alpha \rangle \, .
\end{eqnarray}
We employ $\Delta_{\alpha}$ to indicate the existence of different kinematical
thresholds.
For example, in the $X(3872)$ the charged channel ($D^{+} {D}^{*-}$)
lies $8.06\,{\rm MeV}$ above the neutral channel ($D^{0} \bar{D}^{*0}$).
However, the binding energy is referred relative to the neutral channel.
Therefore in this case we take $\Delta_0 = 0$ and $\Delta_C = 8.06\,{\rm MeV}$.

The previous equation can be easily projected into the plane wave basis
$|\vec{k}, \alpha \rangle$ yielding
\begin{eqnarray}
\Psi_B^{\alpha}(\vec{k})
=&& \nonumber \\ - \frac{2\mu_{\alpha}}{k^2 + \gamma_{\alpha}^2} &&
\sum_{\beta}\,\int \frac{d^3\vec{k}\,'}{(2\pi)^3}\,
\langle \vec{k}| V^{\alpha \beta}| \vec{k}\,' \rangle\,
\Psi_B^{\beta}(\vec{k}') \, , 
\end{eqnarray}
with $\Psi_B^{\alpha}(\vec{k}) = \langle \vec{k}, \alpha | \Psi_B \rangle$
and $\gamma_{\alpha}^2 = -2\mu_{\alpha}\,(E_B + \Delta_{\alpha})$,
where we have assumed all the channels to be below threshold
($E_{\alpha} + \Delta_{\alpha} \leq 0$) for simplicity.
The normalization condition for the wave function is given by
\begin{eqnarray}
\sum_{\alpha}\,\int \frac{d^3\vec{k}}{(2\pi)^3}\,
|\Psi^{\alpha}_{B} (\vec{k})|^2 = 1 \, .
\end{eqnarray}

The partial wave projection of the bound state equation can be done
as in the previous cases.
We will directly consider the general case of an interaction containing
a tensor operator, that is
\begin{eqnarray}
\label{eq:bs-eq-full}
\Psi^{\alpha}_{B, l j} (k)
=&& \nonumber \\ - \frac{2\mu_{\alpha}}{k^2 + \gamma_{\alpha}^2} &&
\sum_{\beta,l'}\int \frac{k'^2 dk'}{2\pi^2}\,
\langle k | V^{\alpha \beta}_{l l' j}| k' \rangle 
\Psi^{\beta}_{B, l' j} (k') \, , 
\end{eqnarray}
where the wave function is subjected to the normalization condition
\begin{eqnarray}
\sum_{\alpha, {\{l\}}_p}\,\int \frac{k^2 dk}{2\pi^2}\,
|\Psi^{\alpha}_{B, l} (k)|^2 = 1 \, .
\end{eqnarray}

\section{Contact Theory}
\label{sec:contact}

In this section we propose a pionless effective field theory (EFT)
description of the interaction between the $D$ and $\bar{D}^*$
mesons conforming the $X(3872)$.
The formulation of a suitable EFT requires the existence of a separation 
of scales in the physical system under consideration.
In the particular case of a two heavy meson system and for momenta which
are not able to resolve the finite size of the heavy mesons
($p < 0.5-1\,{\rm GeV}$),
the system can be described in terms of the non-relativistic
$D$ and $\bar{D}^*$ meson fields, the pion field and
the local interactions between these fields,
as far as they are compatible with the known symmetries of the system
like parity, time reversal, rotational invariance and chiral symmetry.
This description can also be applied for the $B$ and $\bar{B}^*$
system.

The purpose of this section is to illustrate in a simple and amenable manner
the general EFT approach followed in the present work.
For this reason, we will intentionally ignore the pion as
an explicit degree of freedom and employ instead
a contact EFT theory at leading order (${\rm LO}$).
If pion exchanges are weak, as happens in the $X(3872)$~\cite{Fleming:2007rp}, 
the omission of
this degree of freedom is not crucial and the effect will be similar
to neglecting the Coulomb interaction between charged heavy mesons.
But if pion exchanges are relevant for the description of the system,
as probably happens in the bottom sector,
the decision of ignoring them will reduce the range of validity of
the present EFT formulation to momenta below the pion mass
($p < 140\,{\rm MeV}$).
The previous means that, if the energy of a $B\bar{B}^*$ bound state
is above $B_{\rm max} \sim {m_{\pi}^2}/{2 \mu_{B\bar{B}^*}} \sim 4\,{\rm MeV}$
(with $m_{\pi}$ the pion mass and $\mu_{B\bar{B}^*}$ the reduced mass),
the predictions of the contact EFT will no longer be reliable.
In such a case, the contact theory should be regarded as a model.

We describe the $X(3872)$ as a bound state of a $D$ and
$\bar{D}^*$ mesons in a positive C-parity configuration
which interact through a momentum- and
energy-independent contact interaction of the type
\begin{eqnarray}
\langle \vec{k} | V_C | \vec{k}\,' \rangle = C_0 \, ,
\end{eqnarray}
which is regularized with a suitable regulator function.
However, the different masses of the neutral and charged meson pairs,
$m_{D^0} + m_{\bar{D}^{*,0}} = 3871.79\,{\rm MeV}$ and
$m_{D^+} + m_{{D}^{*,-}} = 3879.85\,{\rm MeV}$,
plus the low lying nature of the $X(3872)$,
require the treatment of the neutral ($D^0\bar{D}^{*0}$) and
charged ($D^+ {D}^{*-}$) components of
the $X$ bound state as independent and separate channels.
In the neutral-charged basis, the contact interaction can be expressed
as the following $2 \times 2$ matrix
\begin{eqnarray}
\label{eq:C0-contact-mat}
\langle \vec{k} | V_C | \vec{k}\,' \rangle = C^{D\bar{D}^*}_0
\begin{pmatrix}
1 & 1 \\
1 & 1
\end{pmatrix}
\, ,
\end{eqnarray}
where we have assumed that the strength of the contact interaction is
independent on whether we have a pair of neutral
or charged mesons.

For determining the value of the $C_0^{D\bar{D}^*}$ counterterm,
we follow the formulation of Ref.~\cite{Gamermann:2009uq},
where a good description of the branching ratio
${Br(X \to J/\psi \pi^{+} \pi^{-} \pi^0)}/
{Br(X \to J/\psi \pi^{+} \pi^{-})}$
was achieved.
In agreement with the phenomenological model of
Refs.~\cite{Gamermann:2006nm,Gamermann:2007fi},
the previous work assumes that the contact interaction between the $D$
and $\bar{D}^*$ mesons are saturated by the $D$ meson weak decay
constant $f_D$, that is~\footnote{
Notice that Refs.~\cite{Gamermann:2006nm,Gamermann:2007fi} employ a
different normalization for $f_D$, which is related to ours by a
factor of $\sqrt{2}$.}
\begin{eqnarray}
\label{eq:C0-sat}
C_0^{D\bar{D}^*} \simeq - \frac{1}{2 f_D^2} \, .
\end{eqnarray}
In the language of Refs.~\cite{Gamermann:2006nm,Gamermann:2007fi}
the previous condition is equivalent to assuming that the value of
the contact operator is saturated by the exchange of
a heavy vector meson in the t-channel.
In Ref.~\cite{Gamermann:2009uq}, the counterterm $C_0^{D\bar{D}^*}$
is regularized with a cut-off $\Lambda$ in momentum space
and iterated in the Lippmann-Schwinger equation.
The value of the cut-off is determined by fixing the binding energy
of the $D\bar{D}^*$ bound state.
This scheme generates a small isospin violation at short distances
which explains the previous branching ratio.

It should be noticed that the phenomenological model of
Refs.~\cite{Gamermann:2006nm,Gamermann:2007fi} requires
the coupling of the neutral and charged channels to
the strange one, conformed by the $D_s^+ {D}_s^{*-}$ mesons.
In such a case, the contact interaction can be written as
the following $3 \times 3$ matrix
\begin{eqnarray}
\langle \vec{k} | V_C | \vec{k}\,' \rangle = C^{D\bar{D}^*}_0
\begin{pmatrix}
1 & 1 & 1 \\
1 & 1 & 1 \\
1 & 1 & 1
\end{pmatrix}
\, .
\end{eqnarray}
However, the energy of the $X(3872)$ state is about $210\,{\rm MeV}$
below $D_s^+ {D}_s^{*-}$ the threshold.
This means that the wave number of the strange component in the $X(3872)$
is approximately $\gamma_s = 653\,{\rm MeV}$, representing a very short
range contribution to the wave function, and approximately of
the order of the natural hard scale for hadronic processes.
Therefore we can safely ignore this contribution and the related strange
channel in the EFT formulation proposed in this work~\footnote{In particular,
the strange channel probability within the $X(3872)$ is completely
negligible.
The contribution of the strange channel to the binding energy of the
$B\bar{B}^*$ state is also quite small, as we will see.
}.

The formulation of the pionless EFT for the negative C-parity states can
be constructed by considering the low energy limit of the phenomenological
model of Refs.~\cite{Gamermann:2006nm,Gamermann:2007fi},
which predicts a contact interaction identical
to the positive C-parity case
\begin{eqnarray}
\label{eq:contact-c-neg}
{\langle \vec{k} | V_C | \vec{k}\,' \rangle}_{C=-1} = C^{D\bar{D}^*}_0
\begin{pmatrix}
1 & 1 \\
1 & 1
\end{pmatrix}
\, ,
\end{eqnarray}
with $C^{D\bar{D}^*}_0$ given by Eq.~(\ref{eq:C0-sat}).
In principle, the previous potential implies the existence of a negative
C-parity partner of the $X(3872)$.
However, this conclusion depends on whether we can apply the pionless
EFT of Ref.~\cite{Gamermann:2009uq} to the $C = -1$ case
without substantial modifications.
In this regard, the model of Refs.~\cite{Gamermann:2006nm,Gamermann:2007fi}
requires the negative C-parity $D\bar{D}^*$ state to couple
with (among others) the $J/\Psi\,\eta$ and $J/\Psi\,\eta'$ channels.
The first lies $227\,{\rm MeV}$ below the $D_0 \bar{D}_0^*$ threshold
and the second $189\,{\rm MeV}$ above.
This means that the $C=-1$ state can decay into $J/\Psi\,\eta$ with a
center-of-mass momentum of $k_{J/\Psi \eta} = 460\,{\rm MeV}$,
and has a $J/\Psi\,\eta'$ component in the wave function
with wave number $\gamma_{J/\Psi \eta'} = 517\,{\rm MeV}$.
In addition, the $J/\Psi\,\eta'$ contribution to the wave function
has a remarkable tendency to decay into $J/\Psi\,\eta$ due to
$\eta-\eta'$ mixing.
The aforementioned momentum scales are slightly below or of the order of
the typical hard scale of hadronic processes
($\Lambda_0 \simeq 0.5-1.0\,{\rm GeV}$),
and may require the inclusion of the $J/\Psi$, the $\eta$ and the $\eta'$
as explicit degrees of freedom in a pionless EFT applicable
for the negative C-parity state.
If the $J/\Psi$, $\eta$ and $\eta'$ fields are not considered,
the EFT treatment of this channel may not be reliable unless
a cut-off below $k_{J/\Psi \eta} \sim \gamma_{J/\Psi \eta'} \sim 0.5\,{\rm GeV}$
is employed.
Yet there is no reason to believe that the structure of the ${\rm LO}$
counterterm is similar in the negative and positive C-parity sectors.
This observation gets support from the lack of clear experimental evidence
for the existence of a negative C-parity $D\bar{D}^*$ state
in the region where the $X(3872)$ is located.

This suspicion is confirmed by the results of Ref.~\cite{Gamermann:2009fv},
in which the negative C-parity $D\bar{D}^*$ state either (i) disappears,
if the same regulator is used as in the $X(3872)$ state,
or (ii) moves about $\Delta E_{\rm cm} \simeq - i\,26 \,{\rm MeV}$
in the complex plane, if the regulator is modified as to keep the
real part of the energy of the negative C-parity state slightly
below the $D_0 \bar{D_0}^*$ threshold.
The previous figures serve as a demonstration of the limits of an EFT
formulation containing only $D$ and $\bar{D}^*$ mesons
for the $C = -1$ case.
An additional limitation is provided by the fact that the negative
C-parity partner of the $X(3872)$ has not been observed experimentally,
as commented in the previous paragraph.
That is, contrary to the $C=+1$ case, in which the explanation of
the $J/\Psi \omega$ over $J/\Psi \rho$ branching ratio provides
a test of the model employed for saturation,
we have no analogous external check of the model
of Refs.~\cite{Gamermann:2006nm,Gamermann:2007fi}
when $C=-1$.
In this regard, there is no compelling reason to trust the saturation
condition given by Eq.~(\ref{eq:contact-c-neg}).
We will therefore ignore for the moment the negative C-parity case.

The extension of the previous EFT formulation to the bottom sector
is straightforward: we expect the form of the contact interaction
to be identical in the charm and bottom sectors, that is
\begin{eqnarray}
\label{eq::C0-contact-bottom}
{\langle \vec{k} | V_C | \vec{k}\,' \rangle} = C^{B\bar{B}^*}_0
\begin{pmatrix}
1 & 1 \\
1 & 1
\end{pmatrix}
\, .
\end{eqnarray}
In addition, as a consequence of the tiny mass splitting between
the neutral ($B_0\bar{B_0}^*$) and charged ($B^{+} B^{*-}$)
channels, the coupled channel structure of
the contact potential simplifies to
\begin{eqnarray}
{\langle \vec{k} | V_C | \vec{k}\,' \rangle} \simeq 2\,C^{B\bar{B}^*}_0 \, .
\end{eqnarray}
The strange channel ($B_s^{0} \bar{B}_s^{*0}$) lies about $180\,{\rm MeV}$
above the $B_0\bar{B_0}^*$ and $B^{+} B^{*-}$ threshold.
However, as a consequence of the heavier reduced mass of
the $B\bar{B}^*$ system, the related wave number of
a strange component is more short-ranged than
the corresponding one in the charm sector:
assuming a low lying bound state, we  obtain a wave number
of $\gamma_s \simeq 980\,{\rm MeV}$. 
Therefore we can safely ignore this degree of freedom.

The problem in predicting the existence of bound states in the bottom sector
is how to determine the value of the counterterm and the cut-off.
The key observation in this regard is that we can make sensible estimations
of $C_0^{B\bar{B}^*}$ and $\Lambda_B$ from $C_0^{D\bar{D}^*}$ and $\Lambda_D$
by invoking HQS.
In particular, we expect that (i) the saturation condition for the counterterm
between the $B$ and $\bar{B}^*$ mesons is given by
\begin{eqnarray}
C_0^{B\bar{B}^*} \simeq -\frac{1}{2 f_B^2} \, ,
\end{eqnarray} 
in analogy with Eq.~(\ref{eq:C0-sat}), and (ii) the value of the counterterms,
which roughly represents the inverse size of the heavy mesons,
or equivalently, the binding energy between
the light and heavy quark,
is similar in both cases, modulo $1/m_Q$ corrections, which leads to
\begin{eqnarray}
\Lambda_B = \Lambda_X + \mathcal{O} \left( \frac{1}{m_Q} \right) \, ,
\end{eqnarray}
where $m_Q$ is the mass of the heavy quark and $\Lambda_X$
the value of the cut-off in the $D\bar{D}^*$ system.
Provided the previous assumptions hold, it is trivial to determine
whether there exists a bottom counterpart of the $X(3872)$.

\subsection{Description of the $X(3872)$ State}

We describe the $X(3872)$ state as a $D\bar{D}^*$ system with C-parity $C=+1$.
Following Ref.~\cite{Gamermann:2009uq}, we include both the neutral and charged
components, which means that the wave function reads
\begin{eqnarray}
\Psi^X (k) &=& 
\frac{1}{\sqrt{2}}\,
\left[ 
| D^0 \bar{D}^{*0} \rangle - | D^{*0} \bar{D}^0  \rangle 
\right]\,\Psi^{X_0}(k) \nonumber \\
&+& 
\frac{1}{\sqrt{2}}\,
\left[ 
| D^{+} \bar{D}^{*-} \rangle - | D^{*+} \bar{D}^{-} \rangle 
\right]\,\Psi^{X_C}(k) \, , \nonumber \\
\end{eqnarray}
where we are assuming $\Psi^{X_0}$ and $\Psi^{X_C}$ to be S-wave.
We will determine the wave functions of the system by solving a two-channel
bound state equation with the following interaction 
\begin{eqnarray}
\label{eq:V-D-contact}
\langle k | V^{\alpha,\beta}_{\rm C} | k' \rangle = 
f(\frac{k}{\Lambda})\,C^{D\bar{D}^*}_0(\Lambda)\,f(\frac{k'}{\Lambda}) \, ,
\end{eqnarray}
where $\alpha,\beta = 0, C$, depending on whether we are considering the
neutral or charged channel.
This contact interaction is equivalent to Eq.~(\ref{eq:C0-contact-mat}).
In addition, we have regularized the contact potential of 
Eq.~(\ref{eq:V-D-contact}) with a regulator $f(x)$
fulfilling the conditions: (i) $f(x) \to 1$ for $x \ll 1$
and (ii) $f(x) \to 0$ for $x \gg 1$.
For simplicity we will use a sharp cut-off regulator, $f(x) = \theta(1-x)$,
along this work.

For the previous contact potential, the bound state equation reads
\begin{eqnarray}
\label{eq:BSE-X0}
\Psi^{X_0}(k) &=& \frac{2\mu_{X,0}}{k^2 + \gamma_{X,0}^2}\,f(\frac{k}{\Lambda})\,
F({\Lambda}) \, , \\
\label{eq:BSE-XC}
\Psi^{X_C}(k) &=& \frac{2\mu_{X,C}}{k^2 + \gamma_{X,C}^2}\,f(\frac{k}{\Lambda})\,
F({\Lambda}) \, ,
\end{eqnarray}
with $F(\Lambda)$ a function independent of the momentum $k$,
which is given by
\begin{eqnarray}
\label{eq:BSE-F}
F(\Lambda) = -\,C^{D\bar{D}^*}_0(\Lambda)\,&& \int
\frac{q^2 dq}{2\pi^2}\,f(\frac{q}{\Lambda})\, \nonumber \\ &\times&
\left[ \Psi^{X_0}(q) + \Psi^{X_C}(q) \right] \, . 
\end{eqnarray}
We have taken $\mu_{X_0} = 966.6\,{\rm MeV}$, $\mu_{X_C} = 968.7\,{\rm MeV}$
for the reduced mass of the neutral and charged subsystems.
The wave numbers are defined as 
$\gamma_{X_0}^2 = - 2\mu_{X_0}\,(E_X + \Delta_{X_0})$ and
$\gamma_{X_C}^2 = - 2\mu_{X_C}\,(E_X + \Delta_{X_C})$.
Traditionally the $X(3872)$ bound state energy is referred with respect
to the threshold of the neutral component.
In accordance with this prescription, we take $\Delta_{X_0} = 0$ and
$\Delta_{X_C} =  (m_{D^{+}} + m_{D^{*-}}) - (m_{D^0} + m_{D^{*0}}) 
= 8.06\,{\rm MeV}$ for the kinematical thresholds.
We take $B_{X} = - E_{X} = 0.1-0.6\,{\rm MeV}$.

The bound state equation is trivial to solve, yielding the solutions
\begin{eqnarray}
\Psi^{X_0}(k) &=& \mathcal{N}
\frac{2\mu_{X_0}}{k^2 + \gamma_{X_0}^2}\,f(\frac{k}{\Lambda}) \, , \\
\Psi^{X_C}(k) &=& \mathcal{N}
\frac{2\mu_{X_C}}{k^2 + \gamma_{X_C}^2}\,f(\frac{k}{\Lambda}) \, ,
\end{eqnarray}
where $\mathcal{N}$ is a normalization constant which can be determined
with the normalization condition
\begin{eqnarray}
\int \frac{k^2 dk}{2 \pi^2}\,\left[ 
|\Psi^{X_0}(k)|^2 + |\Psi^{X_C}(k)|^2 \right] = 1 \, . 
\end{eqnarray}
The eigenvalue equation, which describes the running of the counterterm
with the cut-off $\Lambda$, is obtained by inserting the explicit
solution to the wave functions inside the bound state equations,
Eqs.~(\ref{eq:BSE-X0}), ~(\ref{eq:BSE-XC})
and ~(\ref{eq:BSE-F}), yielding
\begin{eqnarray}
&& -\frac{1}{C_0^{D\bar{D}^*}(\Lambda)} = \nonumber \\ &&
\,\int\frac{q^2 dq}{2\pi^2}\,f^2(\frac{q}{\Lambda})\,
\left[ \frac{2\mu_{X_0}}{q^2 + \gamma_{X_0}^2} +
\frac{2\mu_{X_C}}{q^2 + \gamma_{X_C}^2} \right] \, .
\label{eq:C0-D-running}
\end{eqnarray}

As previously stated, we follow the approach of Ref.~\cite{Gamermann:2009uq}
in which the counterterm is saturated by $f_D$.
If the binding energy has been fixed, the saturation condition
determines the value of the cut-off 
\begin{eqnarray}
\label{eq:C0-D-saturation}
C_0^{D\bar{D}^*}(\Lambda = \Lambda_X) = - \frac{1}{2 f_D^2} \, ,
\end{eqnarray}
where we take the value $f_D = 210 \pm 10\,{\rm MeV}$, a value compatible
with recent lattice simulations.
The saturation condition can only be fulfilled for a specific value of
the cut-off $\Lambda = \Lambda_X$.
For the range of bound state energies considered in this work,
and a sharp cut-off function, we obtain
\begin{eqnarray}
\Lambda_X = 555^{+47}_{-44}\,{\rm MeV} \, ,
\end{eqnarray}
where the error accounts for the binding energy window 
and the uncertainty in $f_D$~\footnote{
In contrast, Ref.~\cite{Gamermann:2009uq} takes $f_D =
\sqrt{2} \times 165\,{\rm MeV}$ (i.e. $233\,{\rm MeV}$)
and $B_X = 0.1\,{\rm MeV}$, for which the cut-off
$\Lambda_X = 653\,{\rm MeV}$ is obtained.
}.
The cut-off can be physically interpreted as the scale at which
the system starts to resolve the composite nature of the $D$ and
$D^*$ mesons.
That is, the cut-off is related to the size of the heavy mesons.

\subsection{The $B\bar{B}^*$ Bound State}

The description of the theoretical $B\bar{B}^*$ bound state involves
a neutral and a charged component with positive C-parity, $C = +1$, 
that is
\begin{eqnarray}
\Psi^{B\bar{B}^*} (k) &=& 
\frac{1}{\sqrt{2}}\,
\left[ 
| B^0 \bar{B}^{*0} \rangle - |  B^{*0} \bar{B}^0 \rangle 
\right]\,\Psi^{B\bar{B}^*(0)}(k) \nonumber \\
&+& 
\frac{1}{\sqrt{2}}\,
\left[ 
| B^{+} \bar{B}^{*-} \rangle - | B^{*+} \bar{B}^{-} \rangle 
\right]\,\Psi^{B\bar{B}^*(C)}(k) \, . \nonumber \\
\end{eqnarray}
As we will see later, the isospin breaking effects are negligible
in this case, and the neutral and charged wave functions
are approximately equal.
In analogy with the $D\bar{D}^*$ case, the $B\bar{B}^*$ interaction
is taken to be
\begin{eqnarray}
\label{eq:V-B-contact}
\langle k | V^{\alpha,\beta}_{\rm C} | k' \rangle = 
f(\frac{k}{\Lambda})\,C^{B\bar{B}^*}_0(\Lambda)\,f(\frac{k'}{\Lambda}) \, ,
\end{eqnarray}
where $\alpha,\beta = 0,C$.
The bound state equation reads
\begin{eqnarray}
\Psi^{B\bar{B}^*(0)}(k) &=& \frac{2\mu_{B\bar{B}^*(0)}}
{k^2 + \gamma_{B\bar{B}^*(0)}^2}\,f(\frac{k}{\Lambda})\,
F({\Lambda}) \, , \\
\Psi^{B\bar{B}^*(C)}(k) &=& \frac{2\mu_{B\bar{B}^*(C)}}
{k^2 + \gamma_{B\bar{B}^*(C)}^2}\,f(\frac{k}{\Lambda})\,
F({\Lambda}) \, ,
\end{eqnarray}
with $F$ given by
\begin{eqnarray}
F(\Lambda) = &-&\,C^{B\bar{B}^*}_0(\Lambda)\,\int
\frac{q^2 dq}{2\pi^2}\,f(\frac{q}{\Lambda})\, \nonumber \\ &\times&
\left[ \Psi^{B\bar{B}^*(0)}(q) + \Psi^{B\bar{B}^*(C)}(q) \right] \, .
\end{eqnarray}
We take $\mu_{B\bar{B}^*,0} = 2651.1\,{\rm MeV}$ and
$\mu_{B\bar{B}^*,C} = 2651.01\,{\rm MeV}$ for the reduced masses.
The wave numbers are
$\gamma_{B\bar{B}^*(0)}^2 = - 2\mu_{B\bar{B}^*(0)}\,
(E_{B\bar{B}^*} + \Delta_{B\bar{B}^*,0})$ and
$\gamma_{B\bar{B}^*(C)}^2 = - 2\mu_{B\bar{B}^*(C)}\,
(E_{B\bar{B}^*} + \Delta_{B\bar{B}^*(C)})$.
In this case, the kinematical threshold gaps are given by
$\Delta_{B\bar{B}^{*}(0)} = 0$ and
$\Delta_{B\bar{B}^{*}(C)} = (m_{B^+} + m_{B^{*-}})  - (m_{B^0} + m_{B^{*0}}) 
\simeq -0.35\,{\rm MeV}$ (with an error of $0.27\,{\rm MeV}$).

The solution of the bound state equation is analogous to that found
in the $D\bar{D}^*$ system.
Taking into account the size of the threshold gap,
$\Delta_{B\bar{B}^{*}(C)} \simeq 0.35\,{\rm MeV}$,
we can make the approximation
\begin{eqnarray}
\Psi^{B\bar{B}^*(0)}(k) &=& \Psi^{B\bar{B}^*(C)}(k) = \nonumber \\ &&
\frac{1}{\sqrt{2}}\,\Psi^{B\bar{B}^*}(k)\left[ 1 +  
\mathcal{O}\left( \frac{\Delta_{B\bar{B}^*}}{E_{B\bar{B}^*}} \right) \right]\, ,
\end{eqnarray}
where the relative error will be fairly small for bound state energies
above $3-4\,{\rm MeV}$.
The isospin symmetric wave function is given by
\begin{eqnarray}
\Psi^{B\bar{B}^*}(k) = \mathcal{N}
\frac{2\mu_{B\bar{B}^*}}{k^2 + \gamma_{B\bar{B}^*}^2}
\,f(\frac{k}{\Lambda}) \, , 
\end{eqnarray}
where $\mathcal{N}$ is a normalization constant which may be obtained by
\begin{eqnarray}
\int \frac{k^2 dk}{2 \pi^2}\,|\Psi^{B\bar{B}^*}(k)|^2 = 1 \, . 
\end{eqnarray}
In the isospin symmetric limit, the the eigenvalue equation reads
\begin{eqnarray}
-\frac{1}{C^{B\bar{B}^*}_0(\Lambda)} =
2\,\int\frac{q^2 dq}{2\pi^2}\,f^2(\frac{q}{\Lambda})\,
\frac{2\mu_{B\bar{B}^*}}{q^2 + \gamma_{B\bar{B}^*}^2} \, ,
\label{eq:C0-B-running}
\end{eqnarray}
and finally, the saturation condition is 
\begin{eqnarray}
\label{eq:C0-B-saturation}
C^{B\bar{B}^*}_0(\Lambda = \Lambda_B) = - \frac{1}{2 f_B^2} \, ,
\end{eqnarray}
with $f_B = 195 \pm 10 \,{\rm MeV}$.
From HQS we expect the value of the cut-off in the $B\bar{B}^*$ system
to be similar to the $D\bar{D}^*$ system, 
that is $\Lambda_B \simeq \Lambda_X$,
as already mentioned.
This assumption is equivalent to presuming the size of the $D$ and $B$
mesons not to depend strongly 
on the mass of the heavy quark.

In this case the known and unknown parameters are different
than in the $D\bar{D}^*$ system.
Contrary to the $X(3872)$ case, the bound state energy of the theoretical
$B\bar{B}^*$ state is unknown and needs to be determined.
In this regard, we will use the eigenvalue equation
Eq.~(\ref{eq:C0-B-running}), the saturation condition
Eq.~(\ref{eq:C0-B-saturation}) and the similar cut-off assumption
to predict the expected bound state energy of the $B\bar{B}^*$ system.

The calculation has three error sources: 
(i) the bound state energy of the $X(3872)$,
which ranges from $B_X = 0.1-0.6\,{\rm MeV}$,
(ii) the uncertainties in the value of $f_B$, and
(iii) the error in the relation $\Lambda_B \simeq \Lambda_X$.
The third source of error is unknown, but can be estimated from
heavy quark symmetry: we expect the cut-offs to be equal modulo
$1/m_Q$ corrections 
\begin{eqnarray}
\Lambda_B = \Lambda_X + \mathcal{O}\left( \frac{1}{m_Q} \right) \, ,
\end{eqnarray}
being $m_Q$ the heavy quark mass.
The problem is how to evaluate the size of the $1/m_Q$ corrections.
A possible estimation comes from the observation that $\sqrt{m_D} f_D$ and 
$\sqrt{m_B} f_B$ also differ by $1/m_Q$ corrections
\begin{eqnarray}
\sqrt{m_B} f_B = \sqrt{m_D} f_D + \mathcal{O}\left( \frac{1}{m_Q} \right) \, .
\end{eqnarray}
Assuming that the relative size of these corrections is the same for
the constants and the cut-off, we obtain
\begin{eqnarray}
\left| \frac{\Lambda_B - \Lambda_X}{\Lambda_B} \right| \sim
\left| \frac{\sqrt{m_D}\,f_D}{\sqrt{m_B}\,f_B} - 1 \right| \, ,
\end{eqnarray}
which yields a $\sim 10-15\%$ relative error to the naive relation
$\Lambda_B = \Lambda_X$.
Therefore, we take $\Lambda_B = \Lambda_X\,(1 \pm 0.15)$
as a conservative estimate.
For the $X(3872)$, we obtained from the saturation condition
$\Lambda_X = 555^{+47}_{-44}\,{\rm MeV}$,
in which the uncertainties coming from $B_X$ and $f_D$
are already taken into account.
Considering the $1/m_Q$ corrections and quadratic error propagation,
the cut-off window for the $B\bar{B}^*$ raises to
$\Lambda_B = 555^{+96}_{-94}\,{\rm MeV}$,
where the errors are dominated by the $1/m_Q$ term.

For the previous estimations of $f_B$ and $\Lambda_B$, we obtain a bound
state energy of $B = 45^{+24}_{-35}\,{\rm MeV}$.
%
The binding energy is clearly beyond the limits of a contact range theory,
which means that the inclusion of pion exchanges is mandatory from
the EFT viewpoint.
The explicit inclusion of isospin breaking shifts the bound state energy
about $\Delta B = -0.2\,{\rm MeV}$ for the central value $B = 45\,{\rm MeV}$,
much below the error in the determination of the bound state energy,
and is therefore a negligible effect.
The effect of including the strange component in the charm and bottom sectors
is small, generating a shift of $\Delta B = -4\,{\rm MeV}$ of the
$B\bar{B}^*$, again much smaller than other uncertainty sources;
the probability of the strange state in such a case is $P_s \simeq 2-3\%$.

\section{The Role of One Pion Exchange}
\label{sec:OPE}

In this section we consider the role of one pion exchange
in the $D\bar{D}^*$ and $B\bar{B}^*$ systems.
As in the two nucleon system, we expect the long range interaction
between two heavy mesons to be driven by one pion exchange.
From the EFT perspective, the inclusion of multiple pion exchanges,
which are in turn constrained by the requirement of broken chiral symmetry,
extends the range of validity of the effective description up to
momenta of the order of the $\rho$ mass or the binding energy
of the heavy and light quarks conforming the heavy meson.
Both scales are of the same order of magnitude, yielding a breakdown
scale of $\Lambda_0 \sim 0.5-1\,{\rm GeV}$.
In this respect, we expect the EFT formulation to be able to describe
bound states up to a maximum binding energy of
$B_{\rm max} \sim {\Lambda_{0}^2}/{2 \mu_{B\bar{B}^*}} \sim 50-200\,{\rm MeV}$
for the particular case of the $B\bar{B}^*$ system.

As already noticed in Ref.~\cite{Tornqvist:1993ng}, an interesting difference
with respect to the two nucleon case lies in the role of the short range
interaction,
which does not involve a strongly repulsive core in the heavy meson case.
In fact, the short range force is attractive in S-waves,
as can be deduced from the existence of the $X(3872)$
state.
This factor will help the formation of S-wave bound states
in the $B\bar{B}^*$ system.

\subsection{The One Pion Exchange Potential}

\begin{figure}[htb]
\begin{center}
\includegraphics[height=2.75cm]{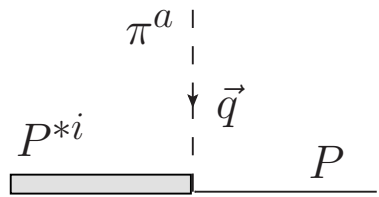}
\end{center}
\caption{Diagram representing the pion-vector-pseudoscalar-meson vertex.
The thin line represents a pseudoscalar meson,
the thick line a vector meson with polarization $i$,
and the dashed line an incoming pion with isospin $a$ and momentum $\vec{q}$.
}
\label{PVpi-vertex}
\end{figure}

\subsubsection{Derivation of the One Pion Exchange Potential}
\label{subsec:OPE-deriv}

The one pion exchange potential between a pseudoscalar (${\rm P}$) and
a vector (${\rm P}^*$) meson (both containing a light quark) can be derived
from the ${\rm PP^*}\pi$ vertex of  Fig.~(\ref{PVpi-vertex}).
In Heavy Hadron Chiral Effective Field Theory~\cite{Wise:1992hn,Yan:1992gz},
the non-relativistic amplitude corresponding to this vertex
in the isospin basis is given by
\begin{eqnarray}
\label{eq:PVpi-vextex}
\mathcal{A}(P^{*i} \to P \pi^{a}) =
\frac{g}{f_{\pi}}\,\frac{\tau^a}{\sqrt{2}}\,
\vec{\epsilon}_i \cdot \vec{q}
\end{eqnarray}
with $\tau_{a}$ the isospin operator (i.e. the Pauli matrices) for the pion
$\pi^{a}$ in the cartesian basis and $\vec{\epsilon}_{i}$ the polarization
vector of the $P^{*i}$ meson.
For the pion weak decay constant, we use the normalization
$f_{\pi} \simeq 130\,{\rm MeV}$.

The value of $g$ represents the coupling of the pion to the light quark
in the pseudoscalar/vector meson.
In non-relativistic quark models, we expect $g = 1$,
while in chiral quark models, $g = 0.75$ (see,
for example, Ref.~\cite{Manohar:1983md}).
However, the best way to determine the value of $g$ is from
the $D^{*}$ decay width.
For the particular cases of the decays ${D^{*}}^{+} \to D^0 \pi^{+}$
and ${D^{*}}^{+} \to D^{+} \pi^{0}$,
the previous amplitude reduces to
\begin{eqnarray} 
\mathcal{A}({D^{*}}^{+} \to D^0 \pi^{+}) &=&
\frac{g}{f_{\pi}}\,\vec{\epsilon}_i \cdot \vec{q} \, ,
\label{eq:Dpiplus-amplitude} \\
\mathcal{A}({D^{*}}^{+} \to D^{+} \pi^{0}) &=&
\frac{g}{\sqrt{2}\,f_{\pi}}\,\vec{\epsilon}_i \cdot \vec{q} \, ,
\label{eq:Dpinot-amplitude}
\end{eqnarray}
as can be trivially checked, yielding the well-known decay rates
\begin{eqnarray}
\Gamma({D^{*}}^{+} \to D^0 \pi^{+}) &=& 
g^2\,\frac{|\vec{q}_{\pi^{+}}|^3}{6 \pi f_{\pi}^2}\, , \\
\Gamma({D^{*}}^{+} \to D^{+} \pi^{0}) &=& 
g^2\,\frac{|\vec{q}_{\pi^{0}}|^3}{12 \pi f_{\pi}^2}\, .
\end{eqnarray}
From these decays (and taking into account the additional electromagnetic
decay ${D^{*}}^{+} \to D^{+}\,\gamma$),
Refs.~\cite{Ahmed:2001xc,Anastassov:2001cw}
obtain the value $g = 0.59 \pm0.01 \pm 0.07$.
For simplicity, we will take $g = 0.6 \pm 0.1$
in the $D\bar{D}^*$ sector.

For the case of the $B\bar{B}^*\pi$ vertex, the value of $g$ cannot
be determined from pion decay.
However, from heavy quark symmetry we should expect a value similar to
that in the $D\bar{D}^*\pi$ vertex.
This expectation seems to be confirmed by lattice
simulations~\cite{Abada:2003un,Negishi:2006sc,Detmold:2007wk,Becirevic:2009yb},
which suggest a value in the range $0.5-0.6$.
We will take $g = 0.55 \pm 0.10$ in the $B\bar{B}^*$ sector.
However there are also other determinations suggesting a smaller
value of $g$.
For example, Ref.~\cite{ElBennich:2010ha} obtains $g = 0.37^{+0.04}_{-0.02}$
based on a study of the Dyson-Schwinger equations.
Therefore, we will also discuss the consequences of having
smaller values of $g$ in the $B\bar{B}^*$ system at
the end of this section.

\begin{figure}[htb]
\begin{center}
\includegraphics[height=4.25cm]{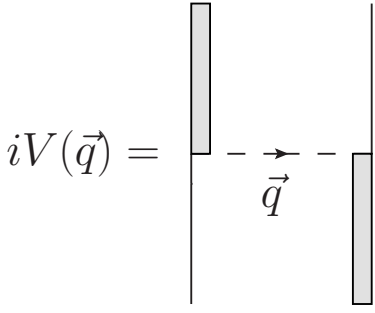}
\end{center}
\caption{Diagram corresponding to the one pion exchange potential between
a pseudoscalar and a vector meson.
}
\label{PV-OPE}
\end{figure}

In the static limit, the one pion exchange potential in momentum space
between a pseudoscalar and vector meson (see Fig.~(\ref{PV-OPE}))
takes the form
\begin{eqnarray}
\label{eq:OPE-p-space}
\tilde{V}_{\rm OPE}(\vec{q}) = - \frac{g^2}{2 f_{\pi}^2} \,
\vec{\tau}_2 \cdot \vec{\tau}_1 \,
\frac{{\vec{\epsilon}_2}^{\,\,*} \cdot \vec{q} \, \vec{\epsilon}_1 \cdot \vec{q}}
{\vec{q^2} + \mu^2 + i \epsilon} \, , 
\end{eqnarray}
where ${\tau}_{1(2)}$ is the isospin operator in vertex $1(2)$,
$\vec{\epsilon}_{1(2)}$ is the $P^*$ meson polarization vector
and $\mu^2 = m_{\pi}^2 - (m_{P^*} - m_P)^2$, with $m_{P^*}$ and $m_P$
the masses of the vector and pseudoscalar mesons.
For the pion propagator we have used
$D_{a b}(q) = \delta_{ab} / (q^2 - m_{\pi}^2 - i \epsilon)$,
where $a$, $b$ are isospin indices.
In the isospin symmetric limit, we will take for the pion mass
the value $m_{\pi} = 138.03\,{\rm MeV}$ (which corresponds to
$(m_{\pi^0} + 2 m_{\pi^{\pm}}) / 3$).
The small imaginary piece ($i \epsilon$) is added for dealing
with the $\mu^2 < 0$ case.

The form of the OPE potential in configuration space can be obtained
by Fourier transforming the momentum space representation of
Eq.~(\ref{eq:OPE-p-space}), in which case we obtain
\begin{eqnarray}
V_{\rm OPE}(\vec{r}) &=&
\int\,\frac{d^3\,q}{(2 \pi)^3}\,\tilde{V}_{\rm OPE}(\vec{q})
\,e^{-i \vec{q} \cdot \vec{r}} \nonumber \\
&=& \frac{g^2}{2 f_{\pi}^2}\,\vec{\tau}_2 \cdot \vec{\tau}_1 \,
({\vec{\epsilon}_2}^{\,\,*} \cdot \vec{\nabla})\,
({\vec{\epsilon}_1} \cdot \vec{\nabla})\,\frac{e^{-\mu r}}{4 \pi r} \, .
\end{eqnarray}
The previous expression can be rewritten as
\begin{eqnarray}
\label{eq:OPE-r-space}
V_{\rm OPE}(\vec{r}) &=& -
\vec{\tau}_2 \cdot \vec{\tau}_1\,
{\vec{\epsilon}_2}^{\,\,*} \cdot {\vec{\epsilon}_1}\,\frac{g^2}{6 f_{\pi}^2}\,
\delta(\vec{r}) \nonumber \\ &+&
\vec{\tau}_2 \cdot \vec{\tau}_1\,\Big[
{\vec{\epsilon}_2}^{\,\,*} \cdot {\vec{\epsilon}_1} W_C (r) \nonumber \\ &+&
(3\,{\vec{\epsilon}_2}^{\,\,*} \cdot \hat{r}\,{\vec{\epsilon}_1} \cdot \hat{r} - 
{\vec{\epsilon}_2}^{\,\,*} \cdot {\vec{\epsilon}_1}) \, W_T (r) \Big] \, ,
\end{eqnarray}
where the central and tensor components of the potential, $W_C$ and $W_T$,
are defined as follows
\begin{eqnarray}
\label{eq:WC}
W_C(r) &=& \frac{g^2 \mu^3}{24 \pi f_{\pi}^2}\,\frac{e^{-\mu r}}{\mu r}\, , \\
\label{eq:WT}
W_T(r) &=& \frac{g^2 \mu^3}{24 \pi f_{\pi}^2}\,\frac{e^{-\mu r}}{\mu r} \, 
\left( 1 + \frac{3}{\mu r} + \frac{3}{\mu^2 r^2} \right) \, .
\end{eqnarray}
It should be noted that by making the replacements $\vec{\epsilon}_i \to 
\vec{\sigma}_i$, $g \to g_A (=1.26)$ and $\mu \to m_{\pi}$ we recover
the one pion exchange potential for the two-nucleon system.
This representation will be useful for determining the channels in which
one pion exchange is more attractive.

A particular problem which arises with the treatment of the one pion exchange
potential in the $D {D}^*$ and $D \bar{D}^*$ systems is that
$\mu^2$ is negative, generating
a singularity in the potential.
This problem is overcome by taking a principal value prescription
\begin{eqnarray}
\frac{1}{\vec{q}^2 - {\tilde{\mu}}^2 + i \epsilon} = \mathcal{P}
\left( \frac{1}{\vec{q}^2 - {\tilde{\mu}}^2} \right)
- \frac{i \pi}{2 q} \delta(q - \tilde{\mu}) \, ,
\end{eqnarray}
where ${\tilde{\mu}}^2 = - \mu^2$ and $\mathcal{P}$
denotes the principal value.
This is equivalent to the change $\mu \to i \tilde{\mu}$
in the coordinate space potentials.
We will ignore the imaginary piece of the potential, which is
related to the decay of the $X(3872)$ to $D\bar{D}\pi$.
A more complete treatment of the $\mu^2 < 0$ case, in which
the $D\bar{D}\pi$ channel is included explicitly,
can be found in Ref.~\cite{Baru:2011rs}.

\subsubsection{C-parity}

We are interested in meson-antimeson systems
where C-parity plays an important role.
In this regard, it should first be noted that the previous OPE potential
has been obtained for a meson-meson system.
The corresponding potential for the meson-antimeson system
in the isospin basis can be obtained by means of
a G-parity transformation, in which case the potential
remains unchanged~\footnote{Of course this depends on the C-parity
convention for the heavy vector meson. In the present work
we have taken $\hat{C} | P^* \rangle = -|\bar{P}^* \rangle$,
which also implies an extra minus sign when performing the G-parity
transformation on the vector meson. If we take into account
the negative G-parity of the pion, the potential in the
isospin basis does not change sign for antiparticles.
If we had chosen the plus sign convention in the C- and G-parity
transformations of the heavy vector mesons, the sign of
the potential would have changed.
However, in such a case we would also have obtained an 
additional minus sign after projecting into well-defined
C-parity states in Eq.~(\ref{eq:pot-G-parity}),
thus leaving the final form of the potential
in the C-parity basis unchanged.}
\begin{eqnarray}
{V}_{\rm OPE}^{P\bar{P}^*}(\vec{r}) &=& V_{\rm OPE}(\vec{r}) \, ,
\end{eqnarray}
where $V_{\rm OPE}(\vec{r})$ is the OPE potential as given
in Eq.~(\ref{eq:OPE-r-space}).

Now we take into account that the previous meson-antimeson potential
is written in the following basis
\begin{eqnarray}
{V}_{\rm OPE}^{P\bar{P}^*}(\vec{r}) &=&
\langle P \bar{P}^*| V | P^* \bar{P}\rangle \nonumber \\
&=& \langle \bar{P} {P}^* | V | \bar{P}^* {P} \rangle \, ,
\end{eqnarray}
where C-parity plays no role.
Therefore, we project the potential into a basis with
well-defined C-parity, that is
\begin{eqnarray}
| P \bar{P}^* (\eta) \rangle &=& \frac{1}{\sqrt{2}}\,
\left[ | P \bar{P}^* \rangle - \eta | P^* \bar{P}  \rangle \right] \, , 
\end{eqnarray}
where $\eta$ represents the intrinsic C-parity of the system~\footnote{
The total C-parity of the heavy meson-antimeson system will be given
by $C = (-1)^{l} \eta$, where $l$ is the orbital angular momentum.}.
In this basis, the potential reads
\begin{eqnarray}
\label{eq:pot-G-parity}
\langle P \bar{P}^* (\eta_{\rm out}) | V | P \bar{P}^* (\eta_{\rm in}) \rangle 
&=& 
- \frac{\eta_{\rm in} + \eta_{\rm out}}{2}\,{V}^{P \bar{P}^*}_{\rm OPE}(\vec{r})
\, ,
\nonumber \\
\end{eqnarray}
that is, the OPE potential conserves C-parity and the sign of the potential
depends on the intrinsic C-parity of the system.

\subsubsection{Isospin Breaking Effects}

In the previous sections we have derived the potential
in the isospin symmetric basis.
However, isospin may be broken due to two effects: (i) the different
masses of the charged and neutral pions and (ii) the different
masses of the charged and neutral heavy mesons.
For the $X(3872)$ (the $D \bar{D}^*$ system), the most important effect
is (ii), due to the weakly bound nature of the $X(3872)$.
On the contrary, for the $B \bar{B}^*$ system we expect both types of isospin
breaking effects to play a relatively minor role. 

To take into account isospin breaking, we consider the charged and neutral
components of the system separately, i.e.
\begin{eqnarray}
| P\bar{P}^*(\eta) \rangle = 
a_0 \,| (P\bar{P}^*(\eta))_0 \rangle + 
a_{\pm}\,| (P\bar{P}^*(\eta))_{\pm} \rangle \, .
\end{eqnarray}
In this neutral-charged basis, we make the substitution~\footnote{
We need to take into account that the $\bar{D}^0$ and $\bar{D}^{*0}$
mesons have an extra minus sign when expressed as isospinors,
that is $\bar{D}^0$ and $\bar{D}^{*0} = -| \frac{1}{2} \frac{1}{2} \rangle$.
}
\begin{eqnarray}
\vec{\tau}_1 \cdot \vec{\tau}_2 \to 
\begin{pmatrix}
- 1 & - 2 \\
- 2 & - 1 
\end{pmatrix} \, ,
\end{eqnarray}
and the OPE potential reads
\begin{eqnarray}
\label{eq:OPE-isospin-breaking}
V^{P\bar{P}^*}_{\rm OPE}(\vec{q}) \to - \eta 
\begin{pmatrix}
- V_{\pi^0}(\vec{q},\mu_0) & - 2 \bar{V}_{\pi^{\pm}}(\vec{q},\mu_{\pm}) \\
- 2 \bar{V}_{\pi^{\pm}}(\vec{q},\mu_{\pm}) & - V_{\pi^0}(\vec{q},\mu_0') 
\end{pmatrix} \, , \nonumber \\ 
\end{eqnarray}
with $V_{\pi}(\vec{q},\mu)$ given by
\begin{eqnarray}
\label{eq:OPE-isospin-breaking-base}
V_{\pi}(\vec{q},\mu) &=& - \frac{g^2}{2 f_{\pi}^2} \,
\frac{{\vec{\epsilon}_2\,}^{*} \cdot \vec{q} \, \vec{\epsilon}_1 \cdot \vec{q}}
{\vec{q^2} + \mu^2} \, , 
\end{eqnarray}
where $\mu$ must be evaluated for the particular channel under consideration.
For the neutral-neutral channel we have
$\mu_0^2 = m_{\pi^0}^2 - (m_{P^{*0}} - m_{P^0})^2$,
while in the charged-charged channel
${\mu_0'}^2 = m_{\pi^0}^2 - (m_{P^{*\pm}} - m_{P^\pm})^2$.
For the term of the potential connecting the charged-neutral channels
there are also two possible definitions for the value of $\mu_{\pm}$,
which are $\mu_{\pm}^2 = m_{\pi^{\pm}}^2 - (m_{P^{*\pm}} - m_{P^0})^2$
and ${\mu_{\pm}'}^2 = m_{\pi^{\pm}}^2 - (m_{P^{*0}} - m_{P^\pm})^2$.
They are different as a consequence of $m_{P^{*\pm}} - m_{P^0} \neq 
m_{P^{*0}} - m_{P^\pm}$.
To include this correction, we simply average over the two possible
definitions of $\mu_{\pm}$
\begin{eqnarray}
\bar{V}_{\pi^{\pm}}(\vec{q},\mu_{\pm}) = \frac{1}{2}\,\left(
V_{\pi^{\pm}}(\vec{q},\mu_{\pm}) + V_{\pi^{\pm}}(\vec{q},\mu_{\pm}')
\right) \, . \nonumber \\ 
\end{eqnarray}

\subsubsection{The Relative Strength of One Pion Exchange}

The strength of the OPE potential in the $P\bar{P}^*$ system
is different in each partial wave (i.e. $^{2S+1}L_J$ channel).
If we ignore the effect of zero-range contributions, bound states between
two heavy mesons will be more probable in channels for which
the long range potential (in this case, OPE) is most attractive.
For this purpose we employ the coordinate space formulation,
which simplifies the identification of the attractive channels.
In addition, to avoid the problems derived from the existence of
a neutral and charged component of the wave function,
which will obscure the analysis, we take the isospin
symmetric limit in the following discussion. 
The methods employed in this subsection were originally developed
in Refs.~\cite{PavonValderrama:2005gu,PavonValderrama:2005wv,PavonValderrama:2005uj}
within the context of the renormalizability of nuclear forces
in chiral EFT.

The configuration space wave function of the heavy meson pair
(with total angular momentum $j$) can be decomposed
into a radial and angular piece
\begin{eqnarray}
\Psi_{B(p j m)}(\vec{r}) = 
\sum_{{\{l\}}_p}\,\frac{u_{l j}(r)}{r} \,\mathcal{Y}_{l j m}(\hat{r}) \, ,
\end{eqnarray}
where $u_{l j}$ are the reduced wave functions of the $P\bar{P}^*$ system and
$\{l\}_p$, $j$, $m$ and $\mathcal{Y}_{l j m}$ were already defined
in Sect.~\ref{sec:tensor_forces}.
Ignoring zero range contributions, the reduced Schr\"odinger equation
for the wave functions $u_{l j}(r)$ reads
\begin{eqnarray}
-u_{l j}''(r) &+& 2\mu_{P {\bar P}^*} \sum_{l'}\,V_{l l' j}(r)\,u_{l' j}(r)
\nonumber \\ &+&
\frac{l(l+1)}{r^2} u_{l j}(r) = - \gamma_{P\bar{P}^*}^2\, u_{l j}(r) \, ,
\end{eqnarray}
where $\mu_{P {\bar P}^*}$ is the reduced mass of the $P\bar{P}^*$ system,
$\gamma_{P\bar{P}^*} = \sqrt{- 2\,\mu_{P {\bar P}^*}\,E_{P\bar{P}^*}}$
the wave number of the bound state, and $E_{P\bar{P}^*}$
the center-of-mass energy of the two meson system.
The partial wave projection of the potential can be obtained from
the following expression
\begin{eqnarray}
V_{l l' j}(r) = &&\sum_{\lambda,\lambda'}\,
\int d\hat{r}\,{{\mathcal Z}^{l \lambda}_{jm}}^{(*)}(\hat{r}) \nonumber \\
&\times&
\langle 1 \lambda | V(\vec{r}) | 1 \lambda' \rangle \,
{\mathcal Z}^{l' \lambda'}_{jm}(\hat{r})\, ,
\end{eqnarray}
as can be checked by adapting the methods of Sect.~\ref{sec:tensor_forces}
to coordinate space.

For the particular case of the OPE potential, $V_{ll' j}$ reads
\begin{eqnarray}
\label{eq:V-r-pw}
V_{l l' j}(r) = - \eta \,\tau\,
\left[ \delta_{l l'} W_C(r) + S_{l l' j} W_T(r) \right] \, ,
\end{eqnarray}
where $\eta$ is the intrinsic C-parity and 
$\tau = \vec{\tau}_2 \cdot \vec{\tau}_1 = 2 I (I+1) - 3$,
with $I$ the total isospin of the two meson system.
$W_C$ and $W_T$ are the central and tensor pieces of the potential,
see Eqs.~(\ref{eq:WC}) and (\ref{eq:WT}).
The $\delta_{l l'}$ and $S_{l l' j}$ represent the partial wave
projection of the ${\vec{\epsilon}_2}^{\,\,*} \cdot {\vec{\epsilon}_1}$ and
 $S_{12}(\hat{r})$ ($ = 3\,{\vec{\epsilon}_2}^{\,\,*} \cdot \hat{r}\,
{\vec{\epsilon}_1} \cdot \hat{r} - {\vec{\epsilon}_2}^{\,\,*}
\cdot {\vec{\epsilon}_1}$) operators respectively.
The $\delta(\vec{r})$ contribution to the OPE potential,
see Eq.~(\ref{eq:OPE-r-space}), has been ignored
as it is a zero range contribution.

The matrix elements of the tensor operator in the partial wave basis are
the following:
(i) for uncoupled channels ($l = j$) we have $S_{j j j} = -1$,
(ii) for coupled channels ($l = j \pm 1$) we obtain
\begin{eqnarray}
{\bf S}_j = \frac{1}{2 j + 1}
\begin{pmatrix}
j-1 & -3 \sqrt{j (j+1)} \\
- 3 \sqrt{j (j+1)} & j+2
\end{pmatrix}
\, ,
\end{eqnarray} 
where the matrix notation stands for $({\bf S}_j)_{l l'} = S_{l l' j}$
~\footnote{As a curious fact,
the matrix elements of the tensor operator for the $P\bar{P}^*$ system
have opposite sign and half the strength of the corresponding matrix
elements in the two-nucleon system.};
finally, (iii) for the $^3P_0$ channel, which is uncoupled
(but for which $l = j+1$),  we have $S_{11 0} = 2$.

Naively we expect the strength of the OPE potential to depend on the interplay
between the central and tensor components.
However, at distances below the pion wavelength ($\mu\,r \leq 1$),
the central piece contribution is small in comparison
with the tensor component~\footnote{
Indeed this is the case in the two nucleon system,
where only the tensor component of the OPE potential needs to be treated
non-perturbatively, while the central piece can always be regarded as
a perturbation.}.
In particular, we expect the $1/r^3$ behaviour of the tensor force 
to determine the relative strength of OPE effects.
At short enough distances, $\mu\,r \ll 1$, we have
\begin{eqnarray}
2\mu_{P {\bar P}^*}\,W_T(r) \to \frac{R_T}{r^3} \, ,
\end{eqnarray}
where $R_T$ is a length scale related to the strength of tensor OPE.
For the OPE potential, we expect in turn
\begin{eqnarray}
2\mu_{P {\bar P}^*}\,V_{l l' j}(r) \to \lambda_{l l' j}\,\frac{R_T}{r^3} \, ,
\end{eqnarray}
where $\lambda_{l l' j} = -\eta\,\tau\,S_{l l' j}$.
That is, the $S_{l l' j}$ matrix determines which channels are attractive.

For the uncoupled channels the identification of the most attractive channels
is trivial as it only requires the multiplication of the matrix element of
the tensor force by $C$ and $\tau$.
For the coupled channels, however, 
the relevant observation is that at short enough distances,
the tensor force ($\sim 1/r^3$) overcomes the centrifugal
barrier ($\sim 1/r^2$) and the central force ($\sim 1/r$),
in which case we can approximate the full Schr\"odinger
equation by
\begin{eqnarray}
-u_{l j}''(r) &-& \eta\,\tau\,\frac{R_T}{r^3}\,\sum_{l'}\,S_{l l' j}\,u_{l' j}(r)
\nonumber \\ &=& - \gamma_{P\bar{P}^*}^2\, u_{l j}(r) +
\mathcal{O}{\left( \frac{u_{l j}}{r^2} \right)}\, ,
\end{eqnarray}
where we have ignored the centrifugal barrier and potential contributions
which diverge less strongly than $1/r^3$.
If we define the following vector, which contains the two components
of the wave function
\begin{eqnarray}
{\bf u}_j = 
\begin{pmatrix}
u_{j-1, j} \\
u_{j+1, j}
\end{pmatrix} \, ,
\end{eqnarray}
we can rewrite the Schr\"odinger equation above as
\begin{eqnarray}
-{\bf u}_{j}''(r) &-& \eta\,\tau\,\frac{R_T}{r^3}\,{\bf S}_j {\bf u}_j
 = - \gamma_{P\bar{P}^*}^2\, {\bf u}_j +
\mathcal{O}{\left( \frac{{\bf u}_{j}}{r^2} \right)} \, .
\end{eqnarray}
In this notation, it is clear that any transformation which diagonalizes
${\bf S}_j$ also diagonalizes the Sch\"odinger equation
at short distances.
This means that the attractive or repulsive character of the tensor
force is determined by the eigenvalues of ${\bf S}_j$, that is
\begin{eqnarray}
{\bf R}_j\,{\bf S}_j\,{\bf R}_j^T =
\begin{pmatrix}
-1 & 0 \\
\phantom{-}0 & 2
\end{pmatrix}
\, ,
\end{eqnarray} 
where ${\bf R}_j$ is the rotation matrix which brings ${\bf S}_j$
into the diagonal basis.
Independently of the values of $\eta$ and $\tau$, there is always an
attractive and a repulsive eigenchannel.
In this regard, at short enough distances the coupled wave function 
contains an attractive component which may generate a bound state.

\begin{table}
\begin{center}
\begin{tabular}{|c|c|c|c|c|c|}
\hline \hline
$I(\eta)$ & 
${}^3S_1-{}^3D_1$ & $^3P_0$ & $^3P_1$ & ${}^3P_2-{}^3F_2$ & $^3D_2$ \\ 
\hline
$0(+1)$ & 
$\{+6,-3\}$ & $+6$ & $-3$ & $\{+6,-3\}$ & $-3$\\
$0(-1)$ & 
$\{-6,+3\}$ & $-6$ & $+3$ & $\{-6,+3\}$ & $+3$ \\
$1(+1)$ & 
$\{-2,+1\}$ & $-2$ & $+1$ & $\{-2,+1\}$ & $+1$\\
$1(-1)$ & 
$\{+2,-1\}$ & $+2$ & $-1$ & $\{+2,-1\}$ & $-1$ \\
\hline \hline
\end{tabular}
\end{center}
\caption{Relative strength of the tensor component of OPE
for the different partial waves in the isospin symmetric basis.
The $I(\eta)$ notation indicates the isospin and the intrinsic C-parity
subchannel under consideration.
Negative values in the table denote channels in which the tensor force
is attractive.
In the coupled channel case, we show the two eigenvalues of
the tensor force.
The strength of tensor OPE in the peripheral uncoupled (coupled) waves
coincides with that of the $^3P_1$ ($^3S_1-{}^3D_1$) channel.
} \label{tab:tensor-strength}
\end{table}

The relative strength of the tensor force for the different partial waves,
that is, the eigenvalues of $\lambda_{l l' j}$, can be found
in Table \ref{tab:tensor-strength}.
The most attractive combination, $\lambda = -6$, is achieved in isoscalar
($I=0$) channels with negative C-parity ($\eta = -1$),
and in particular the uncoupled $^3P_0$ channel
and all the coupled channels ($^3S_1-{}^3D_1$, $^3P_2-{}^3F_2$, etc.).
Next in attractiveness, the isoscalar ($I=0$) channels with positive
C-parity ($\eta = +1$) are to be found, such as the uncoupled $^3P_1$
and the coupled $^3S_1-{}^3D_1$.
However, owing to the repulsive role of the centrifugal barrier,
we only expect the lower partial waves, such as the $^3S_1-{}^3D_1$,
the $^3P_0$ and the $^3P_1$, to be the most promising candidates
for a bound state.

In Table \ref{tab:tensor-cutoff} we list the minimum value of the momentum
space cut-off for having a bound state in the lowest partial waves.
We consider the full OPE potential, as defined in Eqs.~(\ref{eq:OPE-p-space})
or ~(\ref{eq:OPE-r-space}).
That is, we include the zero-range $\delta$ contribution.
As we can see, the previous expectations are approximately fulfilled
expect for the S-waves:
the deviations in the $^3S_1-{}^3D_1$ channel are caused by
the zero range piece of the OPE potential.
In the $^3P_2-{}^3F_2$ channel we do not find bound states for values
of the cut-off below $2\,{\rm GeV}$, probably due to the strong
centrifugal barrier of the F-wave subchannel, ruling out
the possibility of a molecular interpretation in case the
$J^{PC} = 2^{-+}$ assignment~\cite{delAmoSanchez:2010jr}
turns out to be confirmed by future works.

\begin{table}
\begin{center}
\begin{tabular}{|c|c|c|c|c|}
\hline
$D{\bar D}^*$   &\multicolumn{3}{c|}{Partial Wave}  \\
\hline \hline
$I(\eta)$ & ${}^3S_1-{}^3D_1$ & $^3P_0$ & $^3P_1$  \\ \hline
$0(+1)$ &
$840^{+330}_{-200}$ & $-$ & $> 2\,{\rm GeV}$  \\ 
$0(-1)$ & 
$1710^{+710}_{-430}$ & $1050^{+440}_{-260}$ & $-$  \\
$1(+1)$ & 
$> 2\,{\rm GeV}$ & $> 2\,{\rm GeV}$ & $-$  \\ 
$1(-1)$ & 
$> 2\,{\rm GeV}$ & $-$ & $> 2\,{\rm GeV}$  \\ \hline \hline 
$B{\bar B}^*$  &\multicolumn{3}{c|}{Partial Wave}  \\
\hline \hline
$I(\eta)$ & ${}^3S_1-{}^3D_1$ & $^3P_0$ & $^3P_1$  \\ \hline
$0(+1)$ &
$420^{+150}_{-100}$ & $-$ & $940^{+420}_{-240}$  \\
$0(-1)$ & 
$720^{+330}_{-180}$ & $470^{+210}_{-120}$ & $-$  \\
$1(+1)$ & 
$> 2\,{\rm GeV}$ & $1330^{+650}_{-370}$ & $-$  \\
$1(-1)$ & 
$1040^{+470}_{-270}$ & $-$ & $> 2\,{\rm GeV}$ \\ 
\hline \hline
\end{tabular}
\end{center}
\caption{Cut-off (in ${\rm MeV}$) at which the first bound state appears
for the OPE potential in different partial waves.
The $I(\eta)$ column specifies the isospin and the intrinsic C-parity subchannel
(for the $D\bar{D}^*$, where isospin breaking has been taken into
account, $I$ is to be interpreted as "mostly a $I=0/1$ state with
some small admixture of $I=1/0$").
Partial waves without bound states below $\Lambda = 2\,{\rm GeV}$
are not displayed.
Errors take into account the uncertainty in the $P\bar{P}^*\pi$ coupling
$g$.
} \label{tab:tensor-cutoff}
\end{table}

\subsubsection{The Negative C-parity States and the $P^* \bar{P}^*$ System}

\begin{figure}[ttt]
\begin{center}
\includegraphics[height=4.25cm]{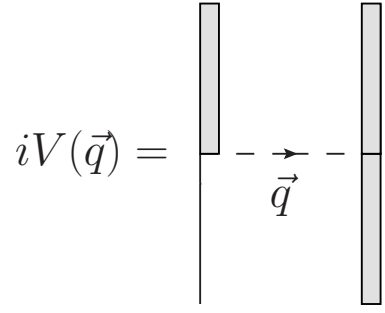}
\end{center}
\caption{One pion exchange diagram mixing the pseudoscalar-vector
($P {P}^*$) and the vector-vector ($P^* {P}^*$) heavy meson systems.}
\label{PV-VV-OPE}
\end{figure}

An interesting feature of the OPE potential, which we have ignored until now,
is that it can mix the $P\bar{P}^*$ and the $P^* \bar{P}^*$
heavy meson systems.
The reason is the following $P^* \bar{P}^* \pi$ vertex
\begin{eqnarray}
\mathcal{A}(P^{*i} \to P^{*j} \pi^a) = 
\frac{g}{f_{\pi}}\frac{\tau_a}{\sqrt{2}}\,i\,
(\vec{\epsilon}_i \times \vec{\epsilon}_j^{\,*}) \cdot \vec{q} \, ,
\end{eqnarray}
which allows the $P {P}^* \to P^* {P}^*$ transition to happen via
the one pion exchange mechanism depicted in Fig.~(\ref{PV-VV-OPE}).
However, owing to the energy separation between the $P\bar{P}^*$
and the $P^* \bar{P}^*$ thresholds, the coupling between
these two systems will not become evident unless
the energy of the $P\bar{P}^*$ bound state is
comparable to the mass difference between
the $P$ and $P^*$ heavy mesons.
In the bottom sector, this requires a binding energy of about $50\,{\rm MeV}$:
for a $B\bar{B}^*$ bound state at threshold, the wave number of
the $B^*\bar{B}^*$ component is $\gamma_{B^*\bar{B}^*} \simeq 0.5\,{\rm GeV}$,
which is of the order of the hard scale of the system
($\Lambda_0 \sim 0.5-1.0\,{\rm GeV}$).
Consequently we can ignore the eventual $B^* \bar{B}^*$ short range
component of the wave function of a $B\bar{B}^*$ bound state
at the price of setting the hard scale at the lower bound
$\Lambda_0 \sim 0.5\,{\rm GeV}$~\footnote{The situation
is in fact very similar to ignoring the $\Delta$ isobar
resonance in the low energy description of the two nucleon system,
where a similar wave number is obtained for the eventual
$N\Delta$ short range component of the deuteron.}.
From the power counting perspective this prescription can be taken
into account by noticing that the mixing between
the $P\bar{P}^*$ and $P^*\bar{P}^*$ channels
is suppressed as the ratio between the $P^*\bar{P}^*$ and $P\bar{P}^*$
propagators near the $P\bar{P}^*$ threshold,
that is, by two orders in the EFT expansion
\begin{eqnarray}
\frac{G_0^{P^* \bar{P}^*}(E)}{G_0^{P\bar{P}^*}(E)}
 = \frac{q^2 - 2\mu_{P\bar{P}^*}\,E}
{q^2 + 2 \mu_{P\bar{P}^*}\,\Delta - 2\mu_{P\bar{P}^*}\,E} \sim 
\left( \frac{P}{\Lambda_0} \right)^2 
\, , \nonumber \\
\end{eqnarray}
where $q \sim \sqrt{2\mu_{P\bar{P}^*}\,E} \sim P$ are considered 
to be light scales,
while $\sqrt{2\mu_{P\bar{P}^*}\,\Delta} \sim \Lambda_0$ is a heavy scale.
In this regard, we expect the mixing of the channels to be an effect
similar in size to chiral two pion exchange (TPE),
which also enters at $\mathcal{O}(P^2)$.

If we are interested in a more detailed account of the effect of
the $P^* \bar{P}^*$ channel in the description of
the $P\bar{P}^*$ bound states, it should be noted that
the conservation of parity, C-parity and total angular momentum
requires that the coupling between the $P\bar{P}^*$ and
$P^* \bar{P}^*$ channels only occurs between states
with the same $J^{PC}$ quantum numbers.
In particular the intrinsic negative C-parity $P\bar{P}^*$ states
will tend to have a stronger mixing with the $P^*\bar{P}^*$ system
than the $\eta = +1$ states.
The reason is that the $^{2S+1}L_J(P\bar{P}^*)$ partial wave
with $\eta = -1$ has the same quantum numbers as
the $^{2S+1}L_J(P^*\bar{P}^*)$ wave~\footnote{We remind that
the C-parity of a $P\bar{P}^*$ and $P^* \bar{P}^*$ systems is 
$C = (-1)^l\,\eta$ and $C = (-1)^{l+s}$ respectively.
This implies that intrinsic negative C-parity $^{2S+1}L_J (P\bar{P}^*)$ states
can always mix with the $^{2S+1}L_J(P^*\bar{P}^*)$ states.
}.
In this regard, the $J^{PC} = 1^{+-}$ $P\bar{P}^*$ state is expected
to have the strongest mixing with the $P^*\bar{P}^*$ system,
as the coupling can happen via S-wave
($^3S_1(P\bar{P}^*) - {}^3S_1 (P^*\bar{P}^*)$).
In the $0^{-+}$ states, the $^3P_0 (P\bar{P}^*)$ and $^3P_0 (P^*\bar{P}^*)$
channels can also mix, although being a P-wave the $^3P_0 (P^*\bar{P}^*)$
component is expected to be suppressed.
The overall effect of this mixing will be attractive:
the short range $P^*\bar{P}^*$ component of the wave function
can always take a configuration which minimizes the energy of
the system, thus increasing the binding.

On the other hand, the intrinsic positive C-parity $P\bar{P}^*$ states
will tend to couple much more weakly with the $P^*\bar{P}^*$ channel.
The extreme case is the $^3P_0 (P\bar{P}^*)$ partial wave ($J^{PC} = 0^{--}$),
which does not contain any $P^*\bar{P}^*$ component as there is
no $P^*\bar{P}^*$ state with such quantum numbers.
In the $1^{++}$ state, the mixing is expected to be small:
the $^3S_1-{}^3D_1 (P\bar{P}^*)$ channel can only couple
with the $^5D_1 (P^*\bar{P}^*)$ partial wave.
In these cases we can probably ignore the $P\bar{P}^*$ mixing 
without significantly reducing the convergence
of the theory.

\subsection{The $X(3872)$ with One Pion Exchange}

We describe the $X(3872)$ as a $^3S_1-{}^3D_1$, $C=+1$ state,
where the charged and neutral components are treated
independently.
The wave function reads
\begin{eqnarray}
\Psi^{X}_{(p j m)} (\vec{k}) &=& 
\frac{1}{\sqrt{2}}\,
\left[ 
| D^0 \bar{D}^{*0} \rangle - | D^{*0} \bar{D}^0 \rangle 
\right] \Psi^{X_0}_{(p j m)}(\vec{k}) \nonumber \\
&+& 
\frac{1}{\sqrt{2}}\,
\left[ 
| D^{+} {D}^{*-} \rangle - |D^{*+} {D}^{-}  \rangle 
\right] \Psi^{X_C}_{(p j m)}(\vec{k}) \, , \nonumber \\
\end{eqnarray}
where $p = +1$, $j = 1$ and $\{l\}_p = 0,2$.
The $\Psi^{X_0}_{(p j m)}(\vec{k})$ and $\Psi^{X_C}_{(p j m)}(\vec{k})$
wave functions can be decomposed into an S-wave and D-wave component
\begin{eqnarray}
\Psi^{X_\alpha}_{(+1 \, 1 m)}(\vec{k}) = \sqrt{4\pi}&\Big[&
\Psi^{X_{\alpha}}_{S}(k) \,\mathcal{Y}_{0 1 m}(\hat{k}) \nonumber \\ &+&
\Psi^{X_{\alpha}}_{D}(k) \,\mathcal{Y}_{2 1 m}(\hat{k}) \Big] \, ,
\end{eqnarray}
where the subscript $\alpha = 0, C$ denotes the neutral and charged components,
and where we have used $S$ and $D$ instead of $l=0,2$ for concreteness.
The normalization condition for the wave function reads
\begin{eqnarray}
\sum_{X_{\alpha}} \, \int \frac{k^2 dk}{2 \pi^2}\,\left[ 
|\Psi^{X_{\alpha}}_S(k)|^2 + |\Psi^{X_\alpha}_D(k)|^2 \right] = 1 \, .
\end{eqnarray}

The wave functions are obtained by solving numerically a four channel
Lippmann-Schwinger equation, Eq.~(\ref{eq:bs-eq-full}),
with the OPE potential and a contact term.
The OPE potential in the neutral-charged basis is taken from 
Eqs.~(\ref{eq:OPE-isospin-breaking})
and ~(\ref{eq:OPE-isospin-breaking-base}).
For $\mu^2 < 0$, we employ the principal value prescription and
ignore the imaginary piece of the OPE potential.
The partial wave projection of the OPE potential can be obtained from
Eq.~(\ref{eq:V_llj}).
For the contact piece, the partial wave projection is trivial,
yielding
\begin{eqnarray}
\langle p, l j | V^{\alpha \beta}_C | p', l' j \rangle = 
C_0^{D\bar{D}^*}\,\delta_{l,0}\delta_{l',0} \, .
\end{eqnarray}
The full potential is regularized with a suitable regulator function $f(x)$
\begin{eqnarray}
\langle p, l j | V^{\alpha \beta} | p', l' j \rangle
\to 
f(\frac{k}{\Lambda})
\langle k, l j | V^{\alpha \beta} | p', l' j \rangle
f(\frac{k'}{\Lambda}) \, , \nonumber \\
\end{eqnarray}
where $\alpha,\beta = 0, C$ and
$V^{\alpha \beta} = V^{\alpha \beta}_{\rm OPE} + V^{\alpha \beta}_C$.
We will employ a sharp cut-off regulator, $f(x) = \theta(1 - x)$.

As in the contact theory case, the value of the cut-off is determined
from the saturation condition, Eq.~(\ref{eq:C0-D-saturation}).
For the central values $g=0.6$, $B_X = 0.35\,{\rm MeV}$
and $f_D = 210\,{\rm MeV}$,
and a sharp cut-off regulator, the saturation cut-off
\begin{eqnarray}
\Lambda_X = 396^{+43}_{-43}\,{\rm MeV} \, 
\end{eqnarray}
is obtained, where the spread in the cut-off value comes from
the uncertainty in $g$, $B_X$ and $f_D$~\footnote{
In particular, we have
$\Lambda_X = 396^{+38}_{-37}\,{}^{+7}_{-10}\,{}^{+19}_{-19}\,{\rm MeV}$,
where the errors come from $g$, $B_X$ and $f_D$ respectively.
For the final value, we have assumed that the errors add quadratically.
}.

\subsubsection{The Negative C-Parity State}

An interesting advantage of including pion exchanges explicitly
is that the EFT framework can be now applied to
the negative C-parity sector.
With OPE the saturation cut-off takes a smaller size than the typical momenta
associate to the $J/\Psi \eta$ and $J/\Psi \eta'$ channels,
which means that we are within the range of applicability
of the theory. 
In this regard, while the contact theory seems to predict a negative C-parity
partner of the $X(3872)$, the inclusion of OPE in the $I(J^{PC}) = 0(1^{+-})$
channel prevents the formation of a bound state unless the cut-off
$\Lambda$ is greater than $760\,{\rm MeV}$.
However $\Lambda_X$ turns out to be too small, meaning that OPE explains
in a natural manner why the negative C-parity state does not exist.

\subsection{$B\bar{B}^*$ Bound States with One Pion Exchange}

The number of bound states between the $B\bar{B}^*$ mesons depends
on the natural value of the cut-off for this system.
Assuming the relation $\Lambda_B \simeq \Lambda_X$,
and taking into account the results from Table \ref{tab:tensor-cutoff},
we expect to find two states in the $B\bar{B}^*$ system:
a $I(J^{PC}) = 0(1^{++})$ and a $I(J^{PC}) = 0(0^{-+})$ state.
The $0(1^{++})$, $^3S_1-{}^3D_1$ state is almost universally predicted in works
investigating $B\bar{B}^*$ molecular states, see for example
Refs.~\cite{AlFiky:2005jd,Liu:2008tn,Lee:2009hy,Ding:2009vj}.
The prediction of a $^3P_0$ resonant state is less usual,
as most works concentrate in the S-wave states.
In addition, owing to the fundamental role of contact interactions
in the formation of heavy meson bound states,
we will find that the negative C-parity $0(1^{+-})$ state is bound too.
In Table~\ref{tab:bs-bottom} there is a summary of the binding energies
of the bound states and resonances predicted in the present work.

\begin{table}
\begin{center}
\begin{tabular}{|c|c|c|c|c|}
\hline \hline
Theory & $^{2s+1}L_J$ & $I(J^{PC})$ & $B$ ($E_{\rm cm}$) & $\Lambda$ \\ \hline
Contact & $^3S_1-{}^3D_1$ & $0(1^{++})$ & $45^{+24}_{-35}$ & $555^{+96}_{-94}$\\ 
\hline
OPE & $^3S_1-{}^3D_1$ & $0(1^{++})$ & $20^{+18}_{-12}$ & $396^{+73}_{-73}$ \\
OPE & $^3S_1-{}^3D_1$ & $0(1^{+-})$ & $6.4^{+7.3}_{-4.6}$ & $396^{+73}_{-73}$ \\
\hline
OPE & $^3P_0$ & $0(0^{-+})$ &
$0.7^{+\infty}_{-1.9} - 1.1^{+\infty}_{-1.6}\frac{i}{2}$ & $396^{+73}_{-73}$  \\
\hline \hline
\end{tabular}
\end{center}
\caption{
Summary of bound states and resonances in the bottom sector.
For the bound states we show the binding energy ($B = - E_{\rm cm}$),
while for the $^3P_0$ resonance we indicate the center of mass energy
of the state. 
} \label{tab:bs-bottom}
\end{table}

\subsubsection{The Positive C-parity $^3S_1-{}^3D_1$ Bound State}

In first place, we consider a theoretical $B\bar{B}^*$ bound state 
in the $^3S_1-{}^3D_1$ channel with positive C-parity, total
isospin $I = 0$ and $J^{PC} = 1^{++}$.
For this system, the wave function reads
\begin{eqnarray}
\Psi^{B\bar{B}^*}_{(p j m)} (\vec{k}) = 
\frac{1}{2}\,
&\Big[&
| B^0 \bar{B}^{*0} \rangle - | B^{*0} \bar{B}^0  \rangle \nonumber \\
&+& 
| B^{+} {B}^{*-} \rangle - | B^{*+} {B}^{-} \rangle 
\Big] \, \Psi_{(p j m)}(\vec{k}) \, , \nonumber \\
\end{eqnarray}
where $p = +1$, $j = 1$, ${\{l\}}_p = 0,2$ and we have assumed
isospin symmetry to hold.
As in the $X(3872)$ case, the wave function can be decomposed
into an S- and D-wave piece
\begin{eqnarray}
\Psi_{(+1 \, 1 m)}(\vec{k}) = \sqrt{4\pi}&\Big[&
\Psi_{S}(k) \,\mathcal{Y}_{0 1 m}(\hat{k}) \nonumber \\ &+&
\Psi_{D}(k) \,\mathcal{Y}_{2 1 m}(\hat{k}) \Big] \, ,
\end{eqnarray}
subjected to the normalization condition
\begin{eqnarray}
\int \frac{k^2 dk}{2 \pi^2}\,\left[ 
|\Psi_S(k)|^2 + |\Psi_D(k)|^2 \right] = 1 \, .
\end{eqnarray}

For the $B\bar{B}^*$ system, as a consequence of isospin symmetry,
we solve a two channel Lippmann-Schwinger equation,
Eq.~(\ref{eq:bs-eq-full}).
The full potential results from adding the OPE contribution, as described
in Eq.~(\ref{eq:OPE-p-space}), and the contact theory contribution,
which in the partial wave basis reads
\begin{eqnarray}
\label{eq:contact-B}
\langle p, l j | V_C | p', l' j \rangle = 
2\,C_0^{B\bar{B}^*}\,\delta_{l,0}\delta_{l',0} \, ,
\end{eqnarray}
where a factor of $2$ needs to be included as we are working
in the isospin symmetric basis. 
We assume the contact interaction to be determined by the saturation condition,
Eq.~(\ref{eq:C0-B-saturation}), where we take $f_B = 195\pm 10\,{\rm MeV}$.
The full, partial wave projected potential is then regulated
with a regulator function $f(x)$,
\begin{eqnarray}
\langle p, l j | V | p', l' j \rangle
\to 
f(\frac{k}{\Lambda})
\langle k, l j | V | p', l' j \rangle
f(\frac{k'}{\Lambda}) \, , \nonumber \\
\end{eqnarray}
with $V = V_{\rm OPE} + V_C$.
As in previous cases, we will employ a sharp cut-off, $f(x) = \theta(1 - x)$.
The cut-off is supposed to be $\Lambda_B = \Lambda_X$,
modulo $1/m_Q$ corrections, which we estimate to be of the order of $15\%$.
This translates into the value 
$\Lambda_B = 396^{+73}_{-73}\,{\rm MeV}$,
where the combined uncertainty is the result of quadratic error propagation.
The cut-off error is dominated by the estimation of the size of
the $1/m_Q$ corrections.

For the values of $g$, $f_B$ and $\Lambda_B$ we are using,
we obtain an estimation of the bound state energy of
$B = 20^{+18}_{-12}\,{\rm MeV}$.
The error estimation is completely dominated by the cut-off uncertainty,
which alone would generate $B = 20^{+18}_{-11}\,{\rm MeV}$.
The other sources of error estimation yield
$B = 20.0^{+3.4}_{-2.7}\,{}^{+2.9}_{-2.4}\,{\rm MeV}$, corresponding to the
$g$ and $f_B$ uncertainty respectively.

It is also interesting to notice that Ref.~\cite{Ding:2009vj} predicts a total
of two $1^{++}$ states in the bottom sector: a shallow bound state just below
the $B\bar{B}^*$ threshold and a deeper one with a binding energy
about $140\,{\rm MeV}$.
The reason for the appearance of the two bound states is a unusually big
value of the $B\bar{B}^*\pi$ coupling.
Within the present framework, and taking $g = 0.55$, a second bound state
does not appear up to $\Lambda_B \simeq 1.2\,{\rm GeV}$:
for this cut-off, there is a low-lying bound state at $B \simeq 0$ and
a second bound state at $B \simeq 400\,{\rm MeV}$.
If we take $g = 0.7$, which is a relatively large value for the $B\bar{B}\pi$
coupling, the appearance of the second bound state happens
at $\Lambda_B \simeq 1\,{\rm GeV}$ with a binding energy
$B \simeq 175\,{\rm MeV}$, not far away from the results
of Ref.~\cite{Ding:2009vj}.
From the aforementioned values the possibility of a second bound state
in the $^3S_1-{}^3D_1$ channel seems quite unlikely, in view
of the values of the cut-off obtained in the present work.
However, the required cut-off and the position of a second bound state
may depend on higher order contributions to the potential.

\subsubsection{The Negative C-parity $^3S_1-{}^3D_1$ Bound State}

In second place we consider the negative C-parity partner
of the previous state, that is, a $B\bar{B}^*$ bound state 
in the $^3S_1-{}^3D_1$ channel with total
isospin $I = 0$ and $J^{PC} = 1^{+-}$.
The complete wave function reads in this case
\begin{eqnarray}
\Psi^{B\bar{B}^*}_{(p j m)} (\vec{k}) = 
\frac{1}{2}\,
&\Big[&
| B^0 \bar{B}^{*0} \rangle + | B^{*0} \bar{B}^0 \rangle \nonumber \\
&+& 
| B^{+} {B}^{*-} \rangle + | B^{*+} {B}^{-}  \rangle 
\Big] \, \Psi_{(p j m)}(\vec{k}) \, , \nonumber \\
\end{eqnarray}
where $p = +1$, $j = 1$ and ${\{l\}}_p = 0,2$.
The spatial wave function, $\Psi_{(p j m)}(\vec{k})$, is normalized and
decomposed into an S- and D-wave component in exactly the same way
as its positive C-parity partner.
The short range potential takes the form given by Eq.~(\ref{eq:contact-B})
and is subjected to the saturation condition,
see Eq.~(\ref{eq:C0-B-saturation}),
with $f_B = 195\pm 10\,{\rm MeV}$.
The OPE potential has a overall minus sign with respect to
the $C = +1$ case.

If we employ a sharp cut-off regulator with
$\Lambda_B = 396^{+73}_{-73}\,{\rm MeV}$,
we obtain a binding energy of $B = 6.4^{+7.3}_{-4.6}\,{\rm MeV}$,
where the error is again dominated by the $1/m_Q$ corrections
(i.e. the cut-off uncertainty).
However, contrary to the positive C-parity state,
we do not have an external check of the validity of
the saturation condition employed in this channel.
This represent an additional source of error
which cannot be easily estimated and we should therefore expect
a much larger uncertainty in the position of this state.

\subsubsection{The $^3P_0$ Resonant State}

The last state we will consider in detail is a isoscalar ($I=0$),
positive C-parity, $^3P_0$ resonant state with $J^{PC} = 0^{-+}$.
For this state the wave function can be written as
\begin{eqnarray}
\Psi^{B\bar{B}^*}_{(p j m)} (\vec{k}) = 
\frac{1}{2}\,
&\Big[&
| B^0 \bar{B}^{*0} \rangle + | B^{*0} \bar{B}^0  \rangle \nonumber \\
&+& 
| B^{+} {B}^{*-} \rangle + | B^{*+} {B}^{-} \rangle 
\Big] \, \Psi_{(p j m)}(\vec{k}) \, , \nonumber \\
\end{eqnarray}
where $p=-1$, ${\{l\}}_p = 1$, $j=0$ and $m=0$.
Even though the intrinsic C-parity of this state is negative,
the total C-parity is positive, as interchange of
the meson and the antimeson will generate an
additional minus sign in the wave function.
The wave function only contains a P-wave component, which means that the
separation of the radial and angular piece is trivial
\begin{eqnarray}
\Psi_{(-1 \, 0 0)}(\vec{k}) = \sqrt{4\pi}\,
\Psi_{P}(k) \,\mathcal{Y}_{1 0 0}(\hat{k}) \, .
\end{eqnarray}
Finally, the normalization condition is
\begin{eqnarray}
\int \frac{k^2 dk}{2 \pi^2}\,|\Psi_P(k)|^2 = 1 \, ,
\end{eqnarray}
which only applies when the state is bound.

For obtaining the resonant (bound) state energy we solve a one channel
Lippmann-Schwinger equation with the OPE potential
in the second (first) Riemann sheet
(see Appendix \ref{app:resonances} for details).
The potential is regulated with a sharp cut-off function, where the cut-off
window is the one determined in the previous section (i.e.
$\Lambda_B = 396^{+73}_{-73}\,{\rm MeV}$).
With this cut-off and $g = 0.55 \pm 0.1$, we obtain a resonant
state energy of
\begin{eqnarray}
E_{\rm cm} = 0.7^{+\infty}_{-1.9} - 1.1^{+\infty}_{-1.6}\,\frac{i}{2}\,{\rm MeV}
\, ,
\end{eqnarray}
where the upper error is to be interpreted as the disappearance of the
resonant state, which eventually happens for the smaller values
of the cut-off.
On the other hand, the lower bound is compatible with a shallow bound state.
The interesting feature of the theoretical $^3P_0$ state is that,
at leading order, it does not depend on the value of a contact operator,
only on the estimation for the natural value of the cut-off.
Although the resonance vanishes for small values of $g$ or of the cut-off,
the existence of this states looks more probable than not.
In addition, owing to the intrinsic negative C-parity of this state,
the wave function can contain an appreciable $B^*\bar{B}^*$ component:
this effect will reduce the energy of the system by a small amount,
helping the formation of a bound state in this channel.

\subsubsection{The Isovector States and the $Z_b(10610)$ Resonance}

The recent discovery by the Belle collaboration of two new resonances,
the $Z_b(10610)$ and $Z_b(10650)$~\cite{Collaboration:2011gj},
which lie a few ${\rm MeV}$ above
the $B\bar{B}^*$ and $B^*\bar{B}^*$ thresholds respectively,
provides two interesting candidates for heavy meson molecular states.
The quantum numbers of the two $Z_b$ states are $I^G(J^P) = 1^+(1^+)$.
In the particular case of the $Z_b(10610)$, it is natural 
to interpret this resonance as a low-lying S-wave $B\bar{B}^*$ state.
Several theoretical works~\cite{Bondar:2011ev,Chen:2011zv,Bugg:2011jr,Zhang:2011jj,Yang:2011rp,Voloshin:2011qa,Sun:2011uh,Cleven:2011gp,Cui:2011fj,Navarra:2011xa}
have appeared recently trying to explain the nature and
properties of these two states.

The S-wave molecular interpretation is however not completely trivial
owing to the location of the $Z_b(10610)$ at $4 \pm 2\,{\rm MeV}$
above the $B\bar{B}^*$ threshold.
Within a potential description this requires an explanation
in terms of a $B\bar{B}^*$ resonance.
In this regard, the existence of a resonant state in non-relativistic
scattering depends on two ingredients, namely (i) a repulsive
potential barrier at long distances and (ii) some sort of
short range attraction.
If we consider the form of the OPE potential in the $I=1$, $C = -1$,
$^3S_1-{}^3D_1$ channel, we can appreciate that the OPE potential
between two S-wave mesons is weakly repulsive and given by
\begin{eqnarray}
V_{\rm ss (j=1)}(r) = W_C(r) \, , 
\end{eqnarray}
see Eq.~(\ref{eq:V-r-pw}).
The shorter range attraction is provided by the S- and D-wave mixing,
which in fact is able to bind the system for large enough values
of the cut-off, see Table \ref{tab:tensor-cutoff}.
In addition, being an intrinsic negative C-parity state,
the $B\bar{B}^*$ and $B^*\bar{B}^*$ channels can mix.
If we take into account that the potential barrier provided by OPE is only
about $1\,{\rm MeV}$ high at $r = 0.5\,{\rm fm}$, which is 
insufficient for reaching the current position of
the $Z_b(10610)$ resonance,
it is apparent that a molecular interpretation of this resonant state
requires a coupled channel approach in which the $B^*\bar{B}^*$
is explicitly included.
In principle the argument above certainly renders difficult the direct
extrapolation of the (purely) contact theories employed to describe
the $X(3872)$ to the newly discovered $Z_b(10610)$, as pion exchanges
and coupled channels are expected to be important.
However, the theoretical analysis of $\Upsilon(5S) \to h_b\pi^+\pi^-$ 
and $h_b(2P)\pi^+\pi^-$ decays of Ref.~\cite{Cleven:2011gp} suggests
that the position of the eventual $B\bar{B}^*$ state corresponding
to the $Z_b$ does not need to coincide with the position
obtained from the Breit-Wigner parametrization employed
in Ref.~\cite{Collaboration:2011gj}.
In particular the $Z_b(10610)$ can be located below threshold,
around $5\,{\rm MeV}$ according to Ref.~\cite{Cleven:2011gp},
reopening the possibility of a bound state interpretation.

In general, isovector heavy meson states are not usually predicted
(or even considered) in potential models,
the reason being that the expected strength of the potential is three times
weaker in isovector than in isoscalar states.
%
From Table \ref{tab:tensor-cutoff} we can see that the $I(J^{PC}) = 1(1^{+-})$
state binds at a cut-off $\Lambda \sim 1\,{\rm GeV}$, which is
definitively bigger than the obtained EFT cut-off,
$\Lambda_B \sim 0.4\,{\rm GeV}$,
or the values of the cut-off which bind the two isoscalar states,
$\Lambda \sim 0.4$ and $0.7\,{\rm GeV}$ respectively~\footnote{
In a related note, the recent exploration of meson exchange
in the $B\bar{B}^*$ and
$B^*\bar{B}^*$ systems of Ref.~\cite{Sun:2011uh} also indicates
that the value of the monopolar cut-off which binds
the $Z_b(10610)$ channel is twice the size of
the required cut-off for binding the $I(J^{PC}) = 0(1^{++})$ state,
in qualitative agreement with the results of Table \ref{tab:tensor-cutoff}.}.
This does not mean however that the current estimation of the size of
the cut-off is incorrect.
A more natural explanation lies in the modification of
the short-range dynamics.
In particular, the saturation condition employed in the present work
requires the contact interaction to be $C_0 = - 1/f_B^2$ in isoscalar
channels and $C_0 = 0$ in the isovectors,
which penalizes the formation of $I = 1$ states.
On the contrary, if we assume a zero energy bound state
in the $Z_b(10610)$ channel, 
the isovector counterterm needs to take the value
$C_0 = (-0.3 \pm 0.1) /f_B^2$,
which although non-zero, is still relatively small in comparison
to the isoscalar counterterm.
Moreover the size of the counterterm is expected to decrease further
if we take into account that (i) the $Z_b(10610)$ is above threshold
(see however the previous discussion)
and (ii) the mixing with the $B^*\bar{B}^*$ channel will provide
additional attraction.
In this regard, the only ingredient to accommodate the $Z_b(10610)$ state
within the present framework is a small correction to the short range
dynamics of the system in the line of
\begin{eqnarray}
\label{eq:C0-contact-modified}
\langle \vec{k} | V_C | \vec{k}\,' \rangle = C^{B\bar{B}^*}_0
\begin{pmatrix}
1+\delta & 1-\delta \\
1-\delta & 1+\delta
\end{pmatrix}
\, ,
\end{eqnarray}
instead of the form given in Eq.~(\ref{eq::C0-contact-bottom}),
with $\delta$ some small number
~\footnote{The model of Gamermann and Oset~\cite{Gamermann:2009fv}
already contains this kind of corrections. However, their size 
may be far too small in the case of the $B\bar{B}^*$ system,
giving $\delta \simeq (m_{\rho} / m_{\Upsilon})^2 \sim 0.006$.}.
This will generate a contribution to the contact interaction
in the isovector channel, $C_0 = - \delta / f_B^2$,
thus providing the missing attraction needed
to generate a resonant (or bound) state in the $Z_b(10610)$ channel.

\subsubsection{Higher Order Corrections}

\begin{table}
\begin{center}
\begin{tabular}{|c|c|c|c|c|c|}
\hline \hline
$I(J^{PC})$  & $B^{(n=1)}_{gc}$ & $B^{(n=2)}_{gc}$ & $B^{(n=3)}_{gc}$ & $B^{(n=4)}_{gc}$ & $B_{\rm sc}$
\\ \hline
\hline
$0(1^{++})$ & $29^{+30}_{-18}$ & $23^{+21}_{-14}$ & $21^{+20}_{-13}$ & $21^{+18}_{-13}$ & $20^{+18}_{-12}$ \\
$0(1^{+-})$ & $8^{+10}_{-6}$  
& $6.7^{+8.1}_{-5.0}$ & $6.5^{+7.7}_{-4.7}$ & $6.5^{+7.4}_{-4.7}$ & $6.4^{+7.3}_{-4.6}$ \\
\hline \hline
Cut-off ($\Lambda_B$) & $620^{+120}_{-120}$ & $517^{+96}_{-97}$ & $478^{+89}_{-88}$ & $458^{+84}_{-84}$ & $396^{+73}_{-73}$ \\ \hline
\end{tabular}
\end{center}
\caption{Dependence of the binding energy of the two $B\bar{B}^*$ 
isoscalar bound states with respect to different regulator choices.
In particular we consider gaussian regulators with $n = 1,2,3,4$, labeled
$B^{(n=1,2,3,4)}_{\rm gc}$ and the sharp cut-off regulator $B_{\rm sc}$.
As can be appreciated the central value of the binding energy can change
moderately from one regulator to another.
However, these variations are smaller than the uncertainty coming from
other sources (in particular the $15\%$ error associated with $1/m_Q$).
} \label{tab:bs-regul}
\end{table}

In the previous calculations of the energies of the $B\bar{B}^*$ bound states
we have taken into account three sources of error: (i) the $B\bar{B}^*\pi$
coupling constant $g$, (ii) the weak decay constant $f_B$ and
(iii) the uncertainty in the size of the $1/m_Q$ corrections
for determining a suitable cut-off window.
An additional error source in the EFT formulation is the size of
the contributions from higher order terms.
From the EFT viewpoint, if an observable is computed at order $P^{\nu}$,
the relative error for this observable is expected to be 
$\mathcal{O}(P^{\nu+1}/\Lambda_0^{\nu+1})$,
where $P$ and $\Lambda_0$ are generic notation for the soft
and high scales of the system.
For a ${\rm LO}$ calculation (i.e. $P^0$), the previous estimation
will yield an expected error of $\mathcal{O}(P/\Lambda_0)$.
However, the EFT with pions and heavy mesons fields does not contain
any correction at order $P^1$, the reason being parity conservation.
Therefore we expect the calculations of the binding energy to be accurate
up to
\begin{eqnarray}
B^{\rm EFT} = B^{(0)} + \mathcal{O}\left( \frac{P^2}{\Lambda_0^2} \right) \, ,
\end{eqnarray}
where $B^{\rm EFT}$ is the full value of the binding energy including all
the EFT corrections and $B^{(0)}$ is the ${\rm LO}$ approximation.
In this context, $P$ can be interpreted as the wave number of the bound state,
that is $P \sim \gamma^{(0)} = \sqrt{2\,\mu_{P\bar{P}^*}\,B^{(0)}}$,
with $\mu_{P\bar{P}^*}$ the reduced mass of
the heavy meson-antimeson system.
This means in particular that the relative error is proportional to the
binding energy; therefore, we can write
\begin{eqnarray}
\label{eq:binding-error}
B^{\rm EFT} = B^{(0)} + \mathcal{O}\left( \frac{B^{(0)}}{B^{\rm max}_0} \right)
\, ,
\end{eqnarray}
where $B^{\rm max}_0$ is the maximum bound state energy which can be described by
the EFT, which corresponds to
\begin{eqnarray}
B^{\rm max}_0 = \frac{\Lambda_0^2}{2 \mu_{P\bar{P}^*}} \, .
\end{eqnarray}
Assuming a breakdown scale in the range $\Lambda_0 = 0.5-1.0\,{\rm GeV}$,
we can estimate $B^{\rm max}_0 = 120-500\,{\rm MeV}$ for the $D\bar{D}^*$ system
and $B^{\rm max}_0 = 45-190\,{\rm MeV}$ for $B\bar{B}^*$.
As we are ignoring the mixing with the $D^*\bar{D}^*$ and $B^*\bar{B}^*$
channels, we expect the true breakdown scale to lie closer to
$\Lambda_0 = 0.5\,{\rm GeV}$ than $1\,{\rm GeV}$.

As we can see, the lower estimation for the breakdown of the EFT description
of the $B\bar{B}^*$ bound states suggests moderate corrections
to the estimation of the binding energy of
the $I(J^{PC}) = 0(1^{++})$ state.
This situation requires further analysis in order to check the reliability
of of the results.
In a standard cut-off EFT formulation, a practical way to estimate the size
of the higher order contributions is to vary the cut-off within
a sensible range~\footnote{See for example Ref.~\cite{Epelbaum:2004fk}
for the application of this idea in the context of nucleon-nucleon scattering}.
The underlying idea is that cut-off uncertainties are a higher order effect.
In this sense, varying the cut-off mimics the effect of including (or excluding)
the higher order contributions.
However, in the non-standard formulation employed in this work, cut-off
variations are used for estimating the size of $1/m_Q$ corrections.
In principle, this may be interpreted as the necessity of going to subleading
orders to explicitly check the size of the higher order corrections.
Nevertheless there is a second way of doing things which is to consider
how the results vary with different regulators.
In particular we can check the effect of using gaussian cut-offs of the type
\begin{eqnarray}
f(\frac{k}{\Lambda}) = e^{-\frac{k^{2n}}{\Lambda^{2n}}} \, ,
\end{eqnarray}
for different values of $n$.
As in the sharp cut-off case, we determine $\Lambda$ by fixing the location
of the $X(3872)$ state.

The results for changing the regulator are shown in Table~\ref{tab:bs-regul},
where we have considered the cases $n = 1,2,3,4$~\footnote{
It should be noticed that $2 n$ must be higher than the order $P^\nu$
at which the computations are done to avoid contamination of
the auxiliary cut-off scale at orders below
that of the EFT calculation.
For the ${\rm LO}$ calculation of the present work, this condition
does not translate into a constraint for the value of $n$.
However, if we go to order $P^2$, we should employ at least $n \geq 2$.
}.
We have only considered the two S-wave isoscalar bound states,
as the calculation of the P-wave resonant state energy is much more involved.
As can be seen, the biggest change happens when the $n=1$ gaussian regulator
is used, for which the central value of the cut-off is raised to
$\Lambda = 620\,{\rm MeV}$ and which generates
a change of $10\,{\rm MeV}$ in the binding energy of the $0(1^{++})$
$^3S_1-{}^3D_1$ state (and $2.5\,{\rm MeV}$ in its negative C-parity partner).
The other regulators generate however a much smaller change in the
results.
If we take the $n=1$ gaussian regulator as an upper bound of the ${\rm LO}$
uncertainties, we can appreciate that the EFT error is a bit smaller
than the other error sources (in particular, the $1/m_Q$ corrections).
In addition, the range in which we expect the binding energy to lie does
not change so much.
The uncertainties follow the expectations of Eq.~(\ref{eq:binding-error}),
that is, the relative error grows with the binding energy.
On the other hand, the $40-50\%$ relative error in the $0(1^{++})$ state
is consistent with the lower estimations of the breakdown scale,
indicating that we are well within the range of validity of
the EFT with pions.

\subsubsection{Further Uncertainties in the $B {B}^*\pi$ Coupling}

As previously mentioned in this section,
a particular problem we encounter when considering the OPE potential
in the bottom sector is the determination of a suitable value of
the $B {B}^*\pi$ coupling.
The value we have employed, $g = 0.55 \pm 0.10$, approximately encompasses
most results from lattice QCD, which usually range
from $g = 0.44$~\cite{Becirevic:2009yb}
to $g = 0.63$~\cite{Detmold:2007wk}~\footnote{It should be noted
that in most of lattice QCD calculations the value of the $B B^* \pi$
coupling is computed in the $m_Q \to \infty$ limit, probably with
the exception of Ref.~\cite{Abada:2003un}, in which $g = 0.58$ is
obtained. In general, we expect $g(m_Q = m_b) > g(m_Q \to \infty)$.}.
However, this value of the $B {B}^*\pi$ coupling does not reflect
all the theoretical uncertainties involved in the determination
of this quantity (see, for example, Refs.~\cite{ElBennich:2010ha,Li:2010rh}
for a compilation of values).
In particular, smaller values of this coupling may be possible,
even in the range $g = 0.3-0.4$, a case which we will
consider here.
A small $B B^* \pi$ coupling translates into a weaker OPE potential,
as its strength scales as $g^2$, and a much weaker chiral TPE
($\propto g^4$), meaning that the higher order corrections
owing to pion exchanges will be strongly suppressed.

If we take $g = 0.37^{+0.04}_{-0.03}$ as a reference value,
which was obtained in Ref.~\cite{ElBennich:2010ha}
from the Dyson-Schwinger equations of QCD,
we predict the binding energies $B(1^{++}) = 16^{+14}_{-9}\,{\rm MeV}$
and $B(1^{+-}) = 9.2^{+9.5}_{-6.0}\,{\rm MeV}$ for the positive and
negative C-parity $^3S_1-{}^3D_1$ isoscalar states respectively.
These new values mostly overlap with the
predictions corresponding to $g = 0.55 \pm 0.10$,
which is consistent with the observation that contact operators
are the dominating mechanism in the formation of S-waves
$B\bar{B}^*$ bound states.
On the contrary, the $^3P_0$ ($0^{-+}$) resonant state disappears,
as it depends crucially on the strength of the OPE potential.
For the isovector $Z_b(10610)$ state, the results
do no significantly change: the contact term still needs to be
of the order of $C_0 \simeq -0.3/f_B^2$ to bind the state.

\section{Conclusions}
\label{sec:discussion}

We have considered the $D\bar{D}^*$ and $B\bar{B}^*$ two meson systems
within the framework of a pionless and a pionfull (or chiral)
cut-off EFT at ${\rm LO}$.
In the charmed sector, the existence of the $X(3872)$ state, together
with the saturation hypothesis for the low energy constant $C_0^{D\bar{D}^*}$,
sets the conditions for the applicability of the EFT formulation employed
in this work and results in a natural value of the cut-off $\Lambda_X$.
The determination of the low energy constant and the value of
the cut-off in the $B\bar{B}^*$ case requires invoking HQS
to overcome the absence of experimental information
for this two body system.
In this regard HQS is able to correlate the charm and bottom sectors,
and, in addition, provides error estimates for the contact term
and the cut-off in the $B\bar{B}^*$ system.
This in turn allows to assign errors to the resulting binding energies
of the possible $B\bar{B}^*$ bound states.

The present framework predicts the existence of three isoscalar $B\bar{B}^*$
states with positive C-parity: two $^3S_1-{}^3D_1$ states with positive and
negative C-parity with a binding energy of $20\,{\rm MeV}$ and
$6\,{\rm MeV}$ respectively, and a $^3P_0$ ($0^{-+}$)
resonant state which lies almost at
the $B\bar{B}^*$ threshold.
The different error sources result in relatively large
uncertainties in the previous estimations.
In addition, we expect moderate corrections from subleading order contributions
to the chiral potential between the two heavy mesons.
However, the higher order corrections are probably smaller than the
current uncertainties stemming from the use of the approximate HQS.
Nevertheless, the existence of the bound states is a conclusion that
will likely remain unchanged: if the binding energy is lowered,
the subleading order corrections are expected to decrease,
resulting in a stabilization of the results.

The applicability of the present approach to negative C-parity states
is not entirely free of problems, one of them being
that there is no direct experimental evidence supporting
the short-range dynamics employed in the present work.
This feature translates into an additional error source in the isocalar
$C=-1$ $^3S_1-{}^3D_1$ bound state which we are not able to estimate.
In addition, the existence of the $Z_b(10610)$ state points out to small but
significant deviations from the form of the short range interaction
employed in the present work.
A second issue with the intrinsic negative C-parity states is that they
tend to couple more strongly with the $B^*\bar{B}^*$ system than their
positive C-parity counterparts, an effect that is expected to
increase the overall attraction of the system and help
the formation of bound states,
specifically the aforementioned $Z_b(10610)$.

The extension of the present framework to higher orders in the EFT formulation,
or to other heavy meson systems, such as $P\bar{P}$ and $P^*\bar{P}^*$,
is left for future research.

\begin{acknowledgments}

We would like to thank D. Gamermann and R. Molina
for discussions.
This work was supported by the DGI under contracts FIS2006-03438
and FIS2008-01143, the Generalitat Valenciana contract PROMETEO/2009/0090,
the Spanish Ingenio-Consolider 2010 Program CPAN
(CSD2007-00042) and the EU Research Infrastructure Integrating Initiative
HadronPhysics2.

\end{acknowledgments}

\appendix
\section{Resonant State Equation}
\label{app:resonances}

In this section we extend the Lippmann-Schwinger equation to the second
Riemann sheet for finding resonant and virtual states.
For simplicity we will work in the partial wave decomposition corresponding
to the central potential case; the extension to coupled channels and
tensor forces is straightforward.
The starting point is to use the vertex function,
instead of the more usual wave function.
The vertex function is related with the residue of the T-matrix at
the pole energy, that is
\begin{eqnarray}
\lim_{E \to E_B} \, (E-E_B)
\langle k | T_l(E) | k' \rangle = {\phi_{B,l}(k)\,\phi_{B,l}(k')} \, ,
\nonumber \\
\end{eqnarray}
consequently the relation between the vertex and the wave function is given by
\begin{eqnarray}
\Psi_B(k) = G_0 \phi_B(k) \, .
\end{eqnarray}
Inserting this relationship into the bound state equation,
Eq.~(\ref{eq:bs-central}), and extending the equation to
arbitrary energies, we obtain 
\begin{eqnarray}
\label{eq:vs-central}
\phi_{B, l} (k)
=&& \nonumber \\ - \frac{\mu}{\pi^2} &&
\int \frac{k'^2 dk'}{k'^2 - q^2}\,
\langle k | V_{l}(q)| k' \rangle \phi_{B, l} (k') \, , 
\end{eqnarray}
where $q^2 = -\gamma^2 = 2\mu\,E$.
The advantage of the vertex equation is that it contains
the resolvent operator within the integral, which makes
it possible to select the Riemann sheet by deforming
the integration contour around the $k^2 = q^2$
singularity, see Ref.~\cite{Elster:1997hp}.
The vertex equation only has solutions for the energies
at which the T-matrix has a pole.

In principle, to find the position of resonant and virtual states
it is enough to analytically extend the solutions of the vertex
equation to the second Riemann sheet.
However, in numerical calculations the previous is not trivial.
The numerical evaluation of the resolvent operator, $G_0(E)$,
always chooses the first Riemann sheet, preventing us from
finding either resonant or virtual states.
The solution is to force the selection of the second Riemann sheet,
for example, by changing the integration contour
as previously commented.
Here we will use instead a more informal derivation for extending
Eq.~(\ref{eq:vs-central}) to the second Riemann sheet.
We will assume that the physical scattering region corresponds to
$E + i \epsilon$, with $E$ real and positive.
If we consider $E - i \epsilon$ instead, the resolvent operator takes the form
\begin{eqnarray}
\frac{1}{E - \frac{k^2}{2 \mu}} = 
\mathcal{P}({\frac{1}{E - \frac{k^2}{2 \mu}}}) + 
i \pi\,\frac{\mu}{q}\,\delta(k - q) \, , 
\end{eqnarray}
where $\mathcal{P}$ denotes that the principal value should be taken and
with $E = \frac{q^2}{2 \mu}$.
However, if we were in the second Riemann sheet, we would need the imaginary
piece of the resolvent operator to be negative instead of positive:
we are moving from $E + i \epsilon$ to $E - i \epsilon$
in a continuous manner, which means that the imaginary piece
should be the same in $E + i \epsilon$ as in $E - i \epsilon$.
A practical solution is to add the imaginary piece directly into the resolvent
operator, that is, in the second Riemann sheet we substitute the original
resolvent operator $G_0^{(I)}$ by a new resolvent operator 
 $G_0^{(II)}$
\begin{eqnarray}
G_0^{(I)}(E) \to
G_0^{(II)} = G_0^{(I)}(E) - i 2 \pi \, \frac{\mu}{q}\,\delta(k - q) \, .
\end{eqnarray}
From the point of view of the bound state equation, the previous changes
amounts to the substitution
\begin{eqnarray}
\frac{1}{{k'}^2 - q^2} \to \frac{1}{{k'}^2 - q^2}
+ i \frac{\pi}{q}\,\delta(k' - q) \, ,
\end{eqnarray}
within the integral in Eq.~(\ref{eq:vs-central}).
This change leads to the following set of equations
\begin{eqnarray}
\label{eq:vs-central-2nd}
\phi_{B, l}^{(II)} (k)
=&& 
- i \frac{\mu q}{\pi} \langle k | V_{l} | q \rangle \phi_{B, l}^{(II)} (q)
\nonumber \\ - \frac{\mu}{\pi^2} &&
\int \frac{k'^2 dk'}{{k'}^2 - q^2}\,
\langle k | V_{l} | k' \rangle \phi_{B, l}^{(II)} (k') \, , \\
\phi_{B, l}^{(II)} (q)
=&& 
- i \frac{\mu q}{\pi} \langle q | V_{l} | q \rangle \phi_{B, l}^{(II)} (q)
\nonumber \\ - \frac{\mu}{\pi^2} &&
\int \frac{k'^2 dk'}{{k'}^2 - q^2}\,
\langle q | V_{l} | k' \rangle \phi_{B, l}^{(II)} (k') \, ,
\end{eqnarray}
which are equivalent to the analogous set of equations
obtained in Ref.~\cite{Elster:1997hp}.
Rearranging the different terms in the previous equation to eliminate
the $\phi_{B, l}^{(II)}(q)$ term, we arrive at
\begin{eqnarray}
\phi^{(II)}_{B, l} (k)
=&& \nonumber \\ - \frac{\mu}{\pi^2} &&
\int \frac{k'^2 dk'}{{k'}^2 - q^2}\,
\langle k | W_{l}(q)| k' \rangle \phi^{(II)}_{B, l} (k') \, , 
\end{eqnarray}
where $W_l(q)$ is defined as
\begin{eqnarray}
\langle k | W_{l}(q)| k' \rangle = \langle k | V_{l} | k' \rangle - i
\frac{\mu q}{\pi} 
\frac{\langle k | V_{l} | q \rangle\,\langle q | V_{l} | k' \rangle
}{1 + i \frac{\mu q}{\pi}\,\langle q | V_{l} | q \rangle} \, .
\nonumber \\
\end{eqnarray}
Depending on the value of $q^2$, that is, of $E$, the previous equation will
look for resonances (${\rm Re}(E) > 0$ and ${\rm Im}(E) < 0$) or
virtual states (${\rm Re}(E) < 0$ and ${\rm Im}(E) = 0$).

%

\end{document}